\documentclass[10pt,twosides,a4paper]{book}
\usepackage[utf8]{inputenc}
\usepackage{bbm}
\usepackage{bm}
\usepackage{physics} % adicionei esse pacote
\usepackage{amssymb}
\usepackage[pdftex]{graphicx}
\usepackage{latexsym}
\usepackage[small]{caption2}
\usepackage{amsmath}
\usepackage[english]{babel}
\usepackage{babel}
\usepackage{amsmath,amsthm}
\usepackage{rotating}
\usepackage{multirow}
\usepackage{graphicx}
\usepackage{tocloft}
\usepackage{bbm}
\usepackage{dsfont}
\usepackage{amsfonts}
\usepackage{accents}
\usepackage{pdfpages}
\usepackage[thinc]{esdiff}
\usepackage{hyperref}
\usepackage[numbers,sort&compress,square]{natbib}
\usepackage{natbib}
\usepackage{pdfsync}
\usepackage{fancyhdr}
\usepackage{scalerel,stackengine}
\usepackage{fancyhdr}
\usepackage{booktabs}
\usepackage{listings} % adicionei esse (cÃ³digo)

\usepackage{algorithm}
\usepackage{algpseudocode}

\renewcommand{\cftchappresnum}{CHAPTER }
\newlength{\mylen}
\fancypagestyle{plain}{}

\usepackage{braket}

\usepackage{bbold}
\usepackage{mathtools}
\usepackage{mathrsfs} 

\input{tcilatex}  

\begin{document} 
\chapter[Alexandre P. Costa, Hebert Souza Rego de Oliveira and Alexandre Dodonov \newline
\emph{Comparison of the standard and dressed-picture master equations for
the quantum Rabi model in the ultrastrong coupling regime}]{Comparison of
the standard and dressed-picture master equations for the quantum Rabi model
in the ultrastrong coupling regime}\footnote{Chapter 5 of the book ``Modern Topics in Mathematical,
Quantum and Statistical Physics: Proceedings of the 2025 CIF-UnB conferences''. DOI: 10.29327/5868499. ISBN: 978-65-5563-767-0.}

\label{chapter5}

\markboth{Standard and dressed-picture master equations in the quantum Rabi
model}{A. P. Costa, H. S. Rego de Oliveira and A. Dodonov}

{\large \textbf{Alexandre P. Costa$^{1}$, Hebert S. Rego de Oliveira$^{1}$}}%
\newline
{\large \textbf{and Alexandre Dodonov$^{1,2,a}$}}

\vspace{3mm}

%\vspace{2mm}

\noindent {$^1$ Institute of Physics, University of Brasília, 70910-900,
Brasília, DF, Brazil}

\noindent {$^2$ International Center of Physics, Institute of Physics,
University of Brasília, 70910-900, Brasília, DF, Brazil} %\vspace{2mm}

\noindent {\texttt{$^{a}$adodonov@unb.br}}

%\vspace{2mm}

\section{Abstract}

The goal of this chapter is to investigate the effects of relaxation and
dephasing on the quantum Rabi model in the ultrastrong coupling regime, and
to provide explicit formulas that enable students and researchers to
implement and numerically solve the resulting nonunitary dynamics from first
principles. The quantum Rabi model constitutes the most fundamental
description of light--matter interaction, describing a single two-level
system coupled to a single mode of a quantized cavity field. The ultrastrong
coupling regime is typically defined by coupling strengths satisfying $%
g\gtrsim 0.1\,\omega $, where $\omega $ denotes the cavity-mode frequency.
In this regime, the standard master equation of Quantum Optics---commonly
referred to as the Gorini--Kossakowski--Sudarshan--Lindblad (GKSL) master
equation---is expected to become inaccurate. The underlying reason is that
strong light--matter interaction hybridizes the bare atom and field states,
so that dissipation can no longer be consistently described in the uncoupled
basis. A more consistent treatment must therefore incorporate this
hybridization directly into the derivation of the dissipative terms. One
such approach is the dressed-picture Markovian master equation derived by
Beaudoin, Gambetta, and Blais, in which the qubit--field interaction is
explicitly included when constructing the system--bath coupling operators.
In this chapter, we numerically solve both the standard GKSL master equation
and the dressed-picture master equation (DME) for various initial field
states, including coherent states, odd Schr\"{o}dinger cat states, squeezed
vacuum states, squeezed coherent states, and thermal states. The coupling
strength is varied in the range $0.05\,\omega $ to $0.8\,\omega $. We also
examine photon generation from the vacuum induced by external time-dependent
modulation of the qubit parameters, as well as multiphoton Rabi oscillations
for an initially excited qubit. Two different reservoir spectral densities
are considered: white noise and Ohmic noise. The differences between the two
master-equation approaches are illustrated through numerical results for
several physical observables, including the qubit excited-state population,
the mean photon number, the Mandel $Q$-factor, the negativity (used as a
measure of entanglement), the subsystem purities, and the photon-number
probability distribution at selected times. As expected, for many of these
quantities the predictions of the standard GKSL master equation differ
substantially from those of the dressed-picture master equation. However, in
certain parameter regimes and for specific observables, the discrepancies
remain comparatively small.

\section{Introduction}

Open quantum systems constitutes the field of Quantum Physics devoted to
understanding how a system of interest interacts with its surrounding
environment and how this interaction leads to dissipation, decoherence, and
noise. Although the combined system+environment composite evolves unitarily
according to the Schr\"{o}dinger equation, the reduced dynamics of the
system alone is generally nonunitary and must be described using density
operators and dynamical maps, often in the form of master equations (MEs).
Central topics in this area include the derivation and the evaluation of the
validity of Markovian and non-Markovian master equations for different
physical implementations and regimes of parameters \cite%
{dodon.scully,dodon.carm,dodon.vogel,dodon.breuer,dodon.orszag,dodon.balle,dodon.vega}%
.

The study of open quantum systems in the ultrastrong coupling (USC) regime
has attracted considerable attention over the past decade, particularly due
to the breakdown of standard quantum-optical approaches when the
light--matter coupling strength becomes a significant fraction of the cavity
frequency \cite{dodon.ultra1,dodon.ultra2,dodon.ultra3,dodon.ultra4}. In the USC regime,
defined by $g\gtrsim 0.1\omega $, the rotating-wave approximation fails and
the eigenstates of the system become strongly hybridized. As a consequence,
dissipative treatments based on the standard
Gorini--Kossakowski--Sudarshan--Lindblad (GKSL) master equation derived in
the bare basis may lead to unphysical predictions \cite%
{dodon.werlang,dodon.simone,dodon.Beaudoin}.

A seminal contribution in this context is the work of \emph{Beaudoin,
Gambetta, and Blais} \cite{dodon.Beaudoin}, who derived a Markovian master
equation in the dressed basis (DME, standing for for \textquotedblleft
dressed master equation\textquotedblright ) of the quantum Rabi Hamiltonian
\cite{dodon.rabi1,dodon.rabi2,dodon.rabi3,dodon.rabi4,dodon.rabi5}. By
incorporating the light--matter interaction directly into the derivation of
the dissipative terms, they demonstrated that the standard GKSL master
equation can incorrectly predict spurious excitations and inaccurate steady
states in the USC regime. Although significantly more realistic than the
GKSL ME (and harder to solve, both analytically and numerically), this
equation relied on the secular (rotating-wave) approximation, so that its
validity is compromised in the dispersive regime and at relatively high
photon numbers in cavity and circuit QED. Shortly afterwards, \cite%
{dodon.agarwal} showed how to derive a dressed master equation valid in the
dispersive regime; \cite{dodon.bamba} provided a systematic microscopic
derivation of the dressed master equation and the quantum Langevin equation;
\cite{dodon.Settineri} derived a generalized DME without performing the
secular and low-photon-number approximations; and \cite{dodon.costa} presented a gauge-invariant dressed master equation. Recently, alternative master
equations valid in the USC regime were presented in \cite%
{dodon.Trushechkin2022,dodon.Lednev2024}. It is worth noting that the
numerically \textquotedblleft exact\textquotedblright\ dynamics of a reduced
system under nonperturbative and non-Markovian system--bath interactions can
be obtained via computationally expensive hierarchical equations of motion
(HEOM), as described in the review \cite{dodon.Yoshitaka}.

Although dressed-state master equations are, in principle, more accurate
than the standard GKSL equation in the ultrastrong coupling regime, they
present a significant practical drawback: analytical solutions are generally
out of reach, and numerical implementations require substantial
computational resources and algorithmic care. In particular, the derivation
and implementation of the dressed master equation involve diagonalization of
the interacting Hamiltonian and the evaluation of transition rates in the
dressed basis, which considerably increases numerical complexity.
Consequently, these equations are not immediately accessible to researchers
entering the field. Even the numerical solution of the standard GKSL master
equation for Hilbert spaces involving on the order of 100 photons already
demands nontrivial computational time and memory.

For this reason, in this chapter we provide a step-by-step tutorial on how
to numerically solve both the GKSL master equation and the dressed master
equation of Beaudoin, Gambetta, and Blais. We then illustrate the typical
dissipative dynamics for a variety of initial states, coupling strengths,
and qubit--field detunings. Our numerical results show that, depending on
the parameter regime and the observable considered, the predictions of the
two master equations may either agree reasonably well or differ
dramatically. We therefore hope that this chapter will serve as a practical
guide for students and researchers performing their own numerical studies,
while our examples help delineate the regimes in which the simpler GKSL
master equation provides an acceptable approximation within a given accuracy
threshold.

This chapter is organized as follows. The GKSL ME is described in Section %
\ref{dodon.GKSL} and DME --- in Section \ref{dodon.DME}. Then we illustrate the typical
behavior for the coherent (Section \ref{dodon.coherent}), odd Schr\"{o}dinger cat
(Section \ref{dodon.cat}), squeezed coherent (\ref{dodon.squeezedc}), squeezed vacuum (%
\ref{dodon.squeezedv}), and thermal (Section \ref{dodon.thermal}) states. Section \ref%
{dodon.multi} shows the results for the multiphoton Rabi oscillations for the
initial excited state of the qubit. In Section \ref{dodon.post} we study the
photon generation from vacuum via external modulation of the qubit
transition frequency combined with the preparation of the cavity field state
via post-selection protocols, showing that the generated cavity field state
display metrological power. Finally, Section \ref{dodon.summary} contains the
summary of this work.

\section{GKSL master equation\label{dodon.GKSL}}

A general master equation for the density operator $\rho $ can be written as%
\begin{equation}
\frac{d\rho }{dt}=-i\left[ H,\rho \right] +L\rho
\end{equation}%
where $H$ is the system Hamiltonian, $\left[ ,\right] $ is the commutator
and $L$ is the Liouvillian superoperator that depends on the particular
master equation and whose form can be quite complicated. The
Gorini--Kossakowski--Sudarshan--Lindblad (GKSL) master equation, also known
as the standard master equation of Quantum Optics and derived in almost
every introductory textbook on the subject \cite%
{dodon.carm,dodon.vogel,dodon.scully,dodon.orszag,dodon.breuer}, reads%
\begin{equation}
L_{GKSL}=\kappa _{0}\left( \bar{n}_{\omega }+1\right) D\left[ a\right]
+\kappa _{0}\bar{n}_{\omega }D\left[ a^{\dagger }\right] +\gamma _{0}\left(
\bar{n}_{\Omega }+1\right) D\left[ \sigma _{-}\right] +\gamma _{0}\bar{n}%
_{\Omega }D\left[ \sigma _{+}\right] +\frac{\gamma _{\phi }}{2}D\left[
\sigma _{z}\right]
\end{equation}%
where $\kappa _{0}$ is the cavity damping (relaxation) rate, $\gamma _{0}$
is the qubit damping rate, $\gamma _{\phi }$ is the qubit pure dephasing
rate, $\omega$ is the cavity frequency, $\Omega$ is the qubit transition frequency and $\bar{n}_{\nu }=\left[ e^{\hbar \nu /k_{B}T}-1\right] ^{-1}$ is the
average thermal photon number for the frequency $\nu $, with $T$ being the reservoir's temperature and $k_{B}$ the Boltzman constant.. Here $a$ is the
cavity field annihilator operator, $a^{\dagger }$ is the creation operator, $%
\sigma _{-}=|g\rangle \langle e|$ is the lowering Pauli operator, $\sigma
_{+}=|e\rangle \langle g|$ is the raising Pauli operator and $\sigma
_{z}=|e\rangle \langle e|-|g\rangle \langle g|$ is the pauli $z$-matrix,
where $|g\rangle $ and $|e\rangle $ denote the qubit ground and excited
states, respectively. The Lindblad dissipator is a superoperator defined as%
\begin{equation}
D\left[ O\right] \rho =O\rho O^{\dagger }-\frac{1}{2}\left( O^{\dagger
}O\rho +\rho O^{\dagger }O\right) \,.
\end{equation}

To solve numerically this master equation for the Rabi Hamiltonian (we set $%
\hbar =1$)%
\begin{equation}
H=\omega n+\Omega |e\rangle \langle e|+gX\sigma _{x}~,
\end{equation}%
where $n=a^{\dagger }a$ is the photon number operator and%
\begin{equation}
~X=a+a^{\dagger }~,~\sigma _{x}=\sigma _{+}+\sigma _{-}~.
\end{equation}%
We expand the density operator in the complete orthonormal basis $\left\{
|g,f_{n}\rangle ,|e,f_{n}\rangle \right\} $ as%
\begin{eqnarray}
\rho &=&\sum_{n,m=1}^{Q_{1}+1}\left[ A_{nm}|g,f_{n-1}\rangle \langle
g,f_{m-1}|+B_{nm}|e,f_{n-1}\rangle \langle e,f_{m-1}|\right.  \notag \\
&&\left. +C_{nm}|g,f_{n-1}\rangle \langle e,f_{m-1}|+C_{mn}^{\ast
}|e,f_{n-1}\rangle \langle g,f_{m-1}|\right] ~,  \label{dodon.bare}
\end{eqnarray}%
where $|f_{m}\rangle $ denotes the $m$-th Fock state of the field, $%
n|f_{m}\rangle =m|f_{m}\rangle $. For numerical calculations, we need to
truncate the Hilbert space at $Q_{1}$ photons, assuming that the probability
of $m>Q_{1}$ photons is identically zero (in the actual case, it should be
smaller than some threshold value $\varepsilon $, say $\varepsilon \lesssim
10^{-8}$ for all times). The expansion (\ref{dodon.bare}) permits one to calculate
the average value of any operator $O$ as $\left\langle O\right\rangle =%
\limfunc{Tr}\left[ \rho O\right] $. In particular, the average values of
operators belonging only to the qubit or to the cavity field can be easily
evaluated using the reduced density matrices $\rho _{q}=\limfunc{Tr}%
_{c}\left( \rho \right) $ or $\rho _{c}=\limfunc{Tr}_{q}\left( \rho \right)
$, respectively, where%
\begin{equation}
\rho _{q}=A|g\rangle \langle g|+B|e\rangle \langle e|+C|g\rangle \langle
e|+C^{\ast }|e\rangle \langle g|
\end{equation}%
\begin{equation*}
\rho _{c}=\sum_{N,M=1}^{Q_{1}+1}D_{NM}|f_{N-1}\rangle \langle f_{M-1}|~
\end{equation*}
and%
\begin{equation}
A=\sum_{N=1}^{Q_{1}+1}A_{NN}~,~B=\sum_{N=1}^{Q_{1}+1}B_{NN}~,~C=%
\sum_{N=1}^{Q_{1}+1}C_{NN}~,~D_{NM}\equiv A_{NM}+B_{NM}~.
\end{equation}%
The purities of the qubit and the cavity field are, respectively,
\begin{eqnarray}
\Pi _{q} &=&\limfunc{Tr}\rho _{q}^{2}=A^{2}+B^{2}+2\left\vert C\right\vert
^{2} \label{dodon.por1}\\
\Pi _{c} &=&\limfunc{Tr}\rho _{c}^{2}=\sum_{N,M=1}^{Q_{1}+1}\left\vert
D_{NM}\right\vert ^{2}~.\label{dodon.por2}
\end{eqnarray}

The negativity \cite{dodon.negati}, whose nonzero values attest entanglement
for any bipartite system, is defined as%
\begin{equation}
\mathcal{N}\left( \rho \right) =\left\vert \sum_{\lambda _{i}<0}\lambda
_{i}\right\vert ~,  \label{dodon.nee}
\end{equation}%
where $\lambda _{i}$ are the eigenvalues of the partial transpose of the
total system density operator $\rho $ with respect to any subsystem, denoted
as $\rho ^{T}$. In our case, writing
\begin{equation}
\rho =\sum_{n,m=1}^{Q_{1}+1}\sum_{i,j=1(g)}^{2(e)}\varrho
_{i,n;j,m}|i,f_{n-1}\rangle \langle j,f_{m-1}|~,
\end{equation}%
where $|i=1\rangle \equiv |g\rangle $ and $|i=2\rangle \equiv |e\rangle $,
the matrix elements of the original and partially transposed density
operator are%
\begin{equation}
\langle i,f_{n-1}|\rho |j,f_{m-1}\rangle =\varrho _{i,n;j,m}~,~\langle
i,f_{n-1}|\rho ^{T}|j,f_{m-1}\rangle =\varrho _{j,n;i,m}~.
\end{equation}%
By diagonalyzing the matrix $\rho ^{T}$, the negativity can be easily found.

From the hermiticity condition $\rho =\rho ^{\dagger }$ we can eliminate the
redundant elements by defining for $n>m$%
\begin{equation}
\func{Re}\left( A_{nm}\right) =\func{Re}\left( A_{mn}\right) ,~\func{Im}%
\left( A_{nm}\right) =-\func{Im}\left( A_{mn}\right) ,~\func{Im}\left(
A_{nn}\right) =0~
\end{equation}%
and similar relations hold for $B_{nm}$. Then, for $1\leq p\leq Q_{1}+1$ and
$p\leq j\leq Q_{1}+1$ we define the $\left( Q_{2}+1\right) ^{2}$-dimensional
vector $\mathbf{Y}$ as%
\begin{eqnarray}
\func{Re}\left( A_{p,j}\right) &=&Y\left[ \left( p-1\right) Q_{1}+j-\frac{%
\left( p-1\right) \left( p-2\right) }{2}\right]  \label{dodon.i1} \\
\func{Im}\left( A_{p,j}\right) &=&Y\left[ F_{1}+\left( p-1\right) Q_{1}+j-1-%
\frac{p\left( p-1\right) }{2}\right] \\
\func{Re}\left( B_{p,j}\right) &=&Y\left[ F_{2}+\left( p-1\right) Q_{1}+j-%
\frac{\left( p-1\right) \left( p-2\right) }{2}\right] \\
\func{Im}\left( B_{p,j}\right) &=&Y\left[ F_{3}+\left( p-1\right) Q_{1}+j-1-%
\frac{p\left( p-1\right) }{2}\right] \\
\func{Re}\left( C_{p,j}\right) &=&Y\left[ F_{4}+\left( p-1\right) \left(
Q_{1}+1\right) +j\right] \\
\func{Im}\left( C_{p,j}\right) &=&Y\left[ F_{5}+\left( p-1\right) \left(
Q_{1}+1\right) +j\right] \,,  \label{dodon.i6}
\end{eqnarray}%
where $Q_{2}\equiv 2Q_{1}+1$ and%
\begin{equation}
F_{1}=\frac{\left( Q_{1}+1\right) \left( Q_{1}+2\right) }{2}~,~F_{2}=\left(
Q_{1}+1\right) ^{2}~,~F_{3}=\left( Q_{1}+1\right) ^{2}+\frac{\left(
Q_{1}+1\right) \left( Q_{1}+2\right) }{2}
\end{equation}%
\begin{equation}
F_{4}=2\left( Q_{1}+1\right) ^{2}~,~F_{5}=3\left( Q_{1}+1\right) ^{2}\,.
\end{equation}

After minor algebra, for the zero absolute temperature one obtains the
system of coupled ordinary differential equations for the coefficients of
the density operator%
\begin{eqnarray}
\dot{A}_{NM} &=&-i\omega \left( N-M\right) A_{NM}  \notag \\
&&-ig\left( \sqrt{N}C_{M,N+1}^{\ast }+\sqrt{N-1}C_{M,N-1}^{\ast }-\sqrt{M-1}%
C_{N,M-1}-\sqrt{M}C_{N,M+1}\right)  \notag \\
&&+\kappa _{0}\sqrt{NM}A_{N+1,M+1}-\frac{\kappa _{0}}{2}\left( N+M-2\right)
A_{NM}+\gamma _{0}B_{NM}  \label{dodon.u1}
\end{eqnarray}%
\begin{eqnarray}
\dot{B}_{NM} &=&-i\omega \left( N-M\right) B_{NM}  \notag \\
&&-ig\left( \sqrt{N}C_{N+1,M}+\sqrt{N-1}C_{N-1,M}-\sqrt{M-1}C_{M-1,N}^{\ast
}-\sqrt{M}C_{M+1,N}^{\ast }\right)  \notag \\
&&+\kappa _{0}\sqrt{NM}B_{N+1,M+1}-\left[ \frac{\kappa _{0}}{2}\left(
N+M-2\right) +\gamma _{0}\right] B_{NM}
\end{eqnarray}%
\begin{eqnarray}
\dot{C}_{NM} &=&-i\left[ \omega \left( N-M\right) -\Omega \right] C_{NM}
\notag \\
&&-ig\left( \sqrt{N}B_{N+1,M}+\sqrt{N-1}B_{N-1,M}-\sqrt{M-1}A_{N,M-1}-\sqrt{M%
}A_{N,M+1}\right)  \notag \\
&&+\kappa _{0}\sqrt{NM}C_{N+1,M+1}-\left[ \frac{\kappa _{0}}{2}\left(
N+M-2\right) +\frac{\gamma _{0}}{2}+\gamma _{\phi }\right] C_{NM}~.
\label{dodon.u3}
\end{eqnarray}%
Expressing the coefficients $A_{NM}$, $B_{NM}$ and $C_{NM}$ as components of
the vector $\mathbf{Y}$, one obtains the complete description of the
dissipative dynamics by solving the system of $\left( Q_{2}+1\right) ^{2}$
ordinary differential equations (ODE) of the form $\mathbf{\dot{Y}=MY}$,
where $\mathbf{M}$ is some time-independent square matrix. In our case, we
use the Runge-Kutta-Verner fifth-order and sixth-order method to obtain $%
\mathbf{Y}\left( t\right) $, from which the density operator is recovered
via relations (\ref{dodon.bare}) and (\ref{dodon.i1}) -- (\ref{dodon.i6}).

\section{Dressed picture master equation\label{dodon.DME}}

The dressed master equation (DME) derived in \cite{dodon.Beaudoin} has the
following Liouvillian at zero absolute temperature%
\begin{equation}
L_{DME}\rho =D\left[ \sum_{j}\Phi _{j}|j\rangle \langle j|\right] \rho
+\sum_{j,k>j}\Gamma ^{jk}D\left[ |j\rangle \langle k|\right] \rho ~,
\label{dodon.air}
\end{equation}%
where%
\begin{equation}
\Phi _{j}=\sqrt{\frac{\gamma _{\phi }\left( 0\right) }{2}}\sigma _{z}^{jj}
\end{equation}%
\begin{equation}
\Gamma ^{jk}=\Gamma _{\phi }^{jk}+\Gamma _{\kappa }^{jk}+\Gamma _{\gamma
}^{jk}
\end{equation}%
\begin{equation}
\Gamma _{\phi }^{jk}=\frac{\gamma _{\phi }\left( \Delta _{kj}\right) }{2}%
\left\vert \sigma _{z}^{jk}\right\vert ^{2}~,~\Gamma _{\kappa }^{jk}=\kappa
\left( \Delta _{kj}\right) \left\vert X^{jk}\right\vert ^{2}~,~\Gamma
_{\gamma }^{jk}=\gamma \left( \Delta _{kj}\right) \left\vert \sigma
_{x}^{jk}\right\vert ^{2}
\end{equation}%
and the matrix elements of any operator $O$ are defined as $O^{nm}=\langle
n|O|m\rangle $. Here, $|n\rangle $ is the eigenstate of $H$ with eigenvalue $%
\Lambda _{n}$, $\Delta _{kj}=\Lambda _{k}-\Lambda _{j}$ is the energy
difference and we label the eigenstates in the increasing order of energy
(i. e., $\Lambda _{k+1}\geq \Lambda _{k}$). Althought the equation (\ref{dodon.air}%
) seems complicated, it is still much simpler than the generalized dressed
master equation \cite{dodon.Settineri,dodon.Trushechkin2022,dodon.Yoshitaka}
or other approaches with fewer underlying approximations \cite%
{dodon.Trushechkin2022,dodon.Yoshitaka}.

The manner we chose to solve numerically the DME was to expand the density
operator in the dressed-states basis as%
\begin{equation}
\rho =\sum_{N=0}^{Q_{2}}\sum_{M=0}^{Q_{2}}r_{N,M}|N\rangle \langle M|
\label{dodon.dres}
\end{equation}%
where $Q_{2}+1$ is the total number of orthonormal states in the truncated
dressed-states basis $\left\{ |0\rangle ,\ldots |Q_{2}+1\rangle \right\} $.
From the hermiticity condition $\rho =\rho ^{\dagger }$, which implies $%
r_{n,m}=r_{m,n}^{\ast }$, we define the components of $\left( Q_{2}+1\right)
^{2}$-dimensional vector $\mathbf{Y}$ via the relations%
\begin{eqnarray}
\func{Re}\left( r_{p,j}\right) &=&Y\left( p\cdot Q_{2}+j+1-p\left(
p-1\right) /2\right) ~\text{for }p\leq j \\
\func{Im}\left( r_{p,j}\right) &=&Y\left( \left( Q_{2}+1\right) \left(
Q_{2}+2\right) /2+p\cdot Q_{2}+j-p\left( p+1\right) /2\right) ~\text{for }p<j
\\
\func{Re}\left( r_{p,j}\right) &=&\func{Re}\left( r_{j,p}\right) ;~\func{Im}%
\left( r_{p,p}\right) =0;~\func{Im}\left( r_{p,j}\right) =-\func{Im}\left(
r_{j,p}\right) ~\text{for }p>j~.
\end{eqnarray}

After straightforward calculations, one obtains the system of coupled ODE:%
\begin{eqnarray*}
\func{Re}\left( \dot{r}_{N,M}\right) &=&\left( \Lambda _{N}-\Lambda
_{M}\right) \func{Im}\left( r_{N,M}\right) -\Theta ^{N,M}\func{Re}\left(
r_{N,M}\right) \\
\func{Im}\left( \dot{r}_{N,M}\right) &=&-\left( \Lambda _{N}-\Lambda
_{M}\right) \func{Re}\left( r_{N,M}\right) -\Theta ^{N,M}\func{Im}\left(
r_{N,M}\right)
\end{eqnarray*}%
for $1 \leq M\leq Q_{2}$ and $N<M$, and
\begin{equation*}
\dot{r}_{N,N}^{r}=\sum_{k=N+1}^{Q_{2}}\Gamma ^{Nk}r_{k,k}^{r}-\Upsilon
^{N,N}r_{N,N}^{r}~,~\text{for }0\leq N\leq Q_{2}~,
\end{equation*}%
where%
\begin{equation}
\Theta ^{N,M}=\frac{\gamma _{\phi }\left( 0\right) }{4}\left( \sigma
_{z}^{NN}-\sigma _{z}^{MM}\right) ^{2}+\Upsilon ^{N,M}
\end{equation}%
\begin{equation}
\Upsilon ^{N,M}=\frac{1}{2}\left( \sum_{j=0}^{N-1}\Gamma
^{jN}+\sum_{j=0}^{M-1}\Gamma ^{jM}\right) ~,~\Upsilon
^{N,N}=\sum_{j=0}^{N-1}\Gamma ^{jN}\,.
\end{equation}%
We see that only the diagonal elements have many terms on the right-hand
side, which turns the system relatively easy to solve numerically after one
determines the dressed states either exactly or numerically.

Here we find the dressed states by diagonalizing the Rabi Hamiltonian in the
complete basis $\left\{ |g,f_{n}\rangle ,|e,f_{n}\rangle \right\} $. So we
write
\begin{equation}
|n\rangle =\sum_{i=1}^{Q_{1}+1}c_{i,n}|g,f_{i-1}\rangle
+\sum_{i=Q_{1}+2}^{Q_{2}+1}c_{i,n}|e,f_{i-\left( Q_{1}+2\right) }\rangle ~%
\text{for}~0\leq n\leq Q_{2}~,  \label{dodon.dse}
\end{equation}%
where $Q_{1}$ is the truncation photon number and $Q_{2}=2Q_{1}+1$. In this
basis, the matrix elements read%
\begin{equation}
\sigma _{z}^{nm}=-\sum_{i=1}^{Q_{1}+1}c_{i,n}^{\ast
}c_{i,m}+\sum_{i=Q_{1}+2}^{Q_{2}+1}c_{i,n}^{\ast }c_{i,m}
\end{equation}%
\begin{equation}
\sigma _{x}^{nm}=\sum_{i=1}^{Q_{1}+1}c_{i,n}^{\ast
}c_{i+Q_{1}+1,m}+\sum_{i=Q_{1}+2}^{Q_{2}+1}c_{i,n}^{\ast }c_{i-Q_{1}-1,m}
\end{equation}%
\begin{eqnarray}
X^{nm} &=&\sum_{i=1}^{Q_{1}}\sqrt{i}c_{i,n}^{\ast
}c_{i+1,m}+\sum_{i=2}^{Q_{1}+1}\sqrt{i-1}c_{i,n}^{\ast }c_{i-1,m}  \notag \\
&&+\sum_{i=Q_{1}+2}^{Q_{2}}\sqrt{i-Q_{1}-1}c_{i,n}^{\ast
}c_{i+1,m}+\sum_{i=Q_{1}+3}^{Q_{2}+1}\sqrt{i-Q_{1}-2}c_{i,n}^{\ast
}c_{i-1,m}~,
\end{eqnarray}%
from which all the dissipative rates $\Gamma _{\kappa }^{jk}$ can be found
provided the spectral densities of the reservoirs are known. In this work we
consider two standard examples of spectral densities (recalling that for the
considered zero-temperature regime, the spectral densities are zero for
negative frequencies: $\gamma \left( \Delta _{kj}\right) =\gamma _{\phi
}\left( \Delta _{kj}\right) =\kappa \left( \Delta _{kj}\right) =0$ for $%
\Delta _{kj}<0$). For the \emph{white noise }\cite{dodon.Beaudoin}, we
assume that $\gamma \left( \Delta _{kj}\right) =\gamma _{0}$, $\gamma _{\phi
}\left( \Delta _{kj}\right) =\gamma _{\phi }$ and $\kappa \left( \Delta
_{kj}\right) =\kappa _{0}$ for $\Delta _{kj}\geq 0$, where $\gamma
_{0},\gamma _{\phi }$ and $\kappa _{0}$ are constant damping rate used in
the GKSL ME. For the ohmic noise \cite%
{dodon.ohm,dodon.Settineri}, we assume $\gamma _{\phi
}\left( \Delta _{kj}\right) =\gamma _{\phi }$ for the pure dephasing, while
for the relaxation processes we postulate%
\begin{equation}
\gamma \left( \Delta _{kj}\right) =\gamma _{0}\frac{\Delta _{kj}}{\Omega }%
\exp \left( -\frac{\Delta _{kj}}{\Omega _{c}}\right) ~,~\kappa \left( \Delta
_{kj}\right) =\kappa _{0}\frac{\Delta _{kj}}{\omega }\exp \left( -\frac{%
\Delta _{kj}}{\omega _{c}}\right) ~,
\end{equation}%
where we set the cut-off frequencies $\Omega _{c}=10\Omega $ and $\omega
_{c}=10\omega $.

By solving the system $\mathbf{\dot{Y}}=\mathbf{MY}$, one obtains all the
coefficients $r_{N,M}$ as function of time. Once they are found, it is
simpler to calculate the average values of observables by rewriting the
density operator in the form of Eq. (\ref{dodon.bare}). From the dressed-state
expansion (\ref{dodon.dse}), one finds%
\begin{equation}
A_{NM}=\sum_{n,m=0}^{Q_{2}}r_{n,m}c_{N,n}c_{M,m}^{\ast }
\end{equation}%
\begin{equation}
B_{NM}=\sum_{n,m=0}^{Q_{2}}r_{n,m}c_{N+Q_{1}+1,n}c_{M+Q_{1}+1,m}^{\ast }
\end{equation}%
\begin{equation}
C_{NM}=\sum_{n,m=0}^{Q_{2}}r_{n,m}c_{N,n}c_{M+Q_{1}+1,m}^{\ast }~.
\end{equation}

The initial conditions for the coefficients $r_{nm}$ are found as follows.
For the initial state with the atom in the ground state,
\begin{equation}
\rho _{0}=|g\rangle \langle g|\otimes
\sum_{N,M=0}^{Q_{1}}p_{N,M}|f_{N}\rangle \langle f_{M}|
\end{equation}%
one obtains
\begin{subequations}
\begin{equation}
r_{n,m}\left( 0\right) =\sum_{i,j=0}^{Q_{1}}p_{i,j}c_{i+1,n}^{\ast
}c_{j+1,m}~.  \label{dodon.beh}
\end{equation}%
On the other hand, if it was initially in the excited state, then
\end{subequations}
\begin{equation}
r_{n,m}\left( 0\right) =\sum_{i,j=0}^{Q_{1}}p_{i,j}c_{i+Q_{1}+2,n}^{\ast
}c_{j+Q_{1}+2,m}~.
\end{equation}%
In this work, we assume that the atom is initially in the ground state.

Finally, for a pure initial cavity state, we write $\rho _{c}\left( 0\right)
=|\psi \left( 0\right) \rangle \langle \psi \left( 0\right) |$, where%
\begin{equation}
|\psi \left( 0\right) \rangle =\sum_{n=0}^{Q_{1}}\chi _{n}|n\rangle ~.
\end{equation}%
In Sections \ref{dodon.coherent} -- \ref{dodon.squeezedv} we will describe how to find the recurrence relations for $\chi _{n}$ for some common
states.

\subsection{DME in the bare basis\label{dodon.newsec}}

When the Rabi Hamiltonian depends explicitly on time and the dissipation is
weak, for accuracy reasons it is advantageous to expand the density operador
in terms of the bare basis, as in Eq. (\ref{dodon.bare}). Then, after
straightforward calculations, one gets for $1\leq N,M\leq Q_{1}+1$%
\begin{eqnarray}
\dot{A}_{NM} &=&\text{unitary part}+\sum_{n,m=1}^{Q_{1}+1}[W_{A,A}^{\left(
N,M,n,m\right) }A_{nm}+W_{A,B}^{\left( N,M,n,m\right) }B_{nm} \\
&&+W_{A,C}^{\left( N,M,n,m\right) }C_{nm}+W_{A,D}^{\left( N,M,n,m\right)
}C_{mn}^{\ast }]  \notag
\end{eqnarray}%
\begin{eqnarray}
\dot{B}_{NM} &=&\text{unitary part}+\sum_{n,m=1}^{Q_{1}+1}[W_{B,A}^{\left(
N,M,n,m\right) }A_{nm}+W_{B,B}^{\left( N,M,n,m\right) }B_{nm} \\
&&+W_{B,C}^{\left( N,M,n,m\right) }C_{nm}+W_{B,D}^{\left( N,M,n,m\right)
}C_{mn}^{\ast }]  \notag
\end{eqnarray}%
\begin{eqnarray}
\dot{C}_{NM} &=&\text{unitary part}+\sum_{n,m=1}^{Q_{1}+1}[W_{C,A}^{\left(
N,M,n,m\right) }A_{nm}+W_{C,B}^{\left( N,M,n,m\right) }B_{nm} \\
&&+W_{C,C}^{\left( N,M,n,m\right) }C_{nm}+W_{C,D}^{\left( N,M,n,m\right)
}C_{mn}^{\ast }]  \notag
\end{eqnarray}%
where the unitary part as the same as in Eqs. (\ref{dodon.u1}) -- (\ref{dodon.u3}) (i.e., setting $\kappa _{0}=\gamma _{0}=\gamma _{\phi }=0$) , while the
dissipative contributions are:%
\begin{eqnarray}
W_{A,A}^{\left( N,M,n,m\right) } &=&\sum_{j,k=0}^{Q_{2}}\Phi _{j}\Phi
_{k}c_{n,j}^{\ast }c_{m,k}c_{N,j}c_{M,k}^{\ast } \\
&&-\delta _{m,M}\frac{1}{2}\sum_{k=0}^{Q_{2}}\Phi
_{k}^{2}c_{N,k}c_{n,k}^{\ast }-\delta _{n,N}\frac{1}{2}\sum_{k=0}^{Q_{2}}%
\Phi _{k}^{2}c_{M,k}^{\ast }c_{m,k} \\
&&+\sum_{j=0}^{Q_{2}-1}\sum_{k=j+1}^{Q_{2}}\Gamma ^{jk}\left[
c_{N,j}c_{M,j}^{\ast }c_{n,k}^{\ast }c_{m,k}-\frac{1}{2}\delta
_{m,M}c_{N,k}c_{n,k}^{\ast }-\frac{1}{2}\delta _{n,N}c_{m,k}c_{M,k}^{\ast }%
\right]  \notag
\end{eqnarray}%
\begin{eqnarray}
W_{A,B}^{\left( N,M,n,m\right) } &=&\sum_{j,k=0}^{Q_{2}}\Phi _{j}\Phi
_{k}c_{n+Q_{1}+1,j}^{\ast }c_{m+Q_{1}+1,k}c_{N,j}c_{M,k}^{\ast } \\
&&+\sum_{j=0}^{Q_{2}-1}\sum_{k=j+1}^{Q_{2}}\Gamma ^{jk}\left[
c_{N,j}c_{M,j}^{\ast }c_{n+Q_{1}+1,k}^{\ast }c_{m+Q_{1}+1,k}\right]  \notag
\end{eqnarray}%
\begin{gather}
W_{A,C}^{\left( N,M,n,m\right) }=\sum_{j,k=0}^{Q_{2}}\Phi _{j}\Phi
_{k}c_{n,j}^{\ast }c_{m+Q_{1}+1,k}c_{N,j}c_{M,k}^{\ast }-\delta _{n,N}\frac{1%
}{2}\sum_{k=0}^{Q_{2}}\Phi _{k}^{2}c_{M,k}^{\ast }c_{m+Q_{1}+1,k}  \notag \\
+\sum_{j=0}^{Q_{2}-1}\sum_{k=j+1}^{Q_{2}}\Gamma ^{jk}\left[
c_{N,j}c_{M,j}^{\ast }c_{n,k}^{\ast }c_{m+Q_{1}+1,k}-\delta _{n,N}\frac{1}{2}%
c_{m+Q_{1}+1,k}c_{M,k}^{\ast }\right]
\end{gather}%
\qquad
\begin{eqnarray}
W_{B,A}^{\left( N,M,n,m\right) } &=&\sum_{j,k=0}^{Q_{2}}\Phi _{j}\Phi
_{k}c_{N+Q_{1}+1,j}c_{M+Q_{1}+1,k}^{\ast }c_{n,j}^{\ast }c_{m,k} \\
&&+\sum_{j=0}^{Q_{2}-1}\sum_{k=j+1}^{Q_{2}}\Gamma ^{jk}\left[
c_{N+Q_{1}+1,j}c_{M+Q_{1}+1,j}^{\ast }c_{n,k}^{\ast }c_{m,k}\right]  \notag
\end{eqnarray}%
\begin{eqnarray}
W_{B,B}^{\left( N,M,n,m\right) } &=&\sum_{j,k=0}^{Q_{2}}\Phi _{j}\Phi
_{k}c_{N+Q_{1}+1,j}c_{M+Q_{1}+1,k}^{\ast }c_{n+Q_{1}+1,j}^{\ast
}c_{m+Q_{1}+1,k} \\
&&-\delta _{m,M}\frac{1}{2}\sum_{k=0}^{Q_{2}}\Phi
_{k}^{2}c_{N+Q_{1}+1,k}c_{n+Q_{1}+1,k}^{\ast }-\delta _{n,N}\frac{1}{2}%
\sum_{k=0}^{Q_{2}}\Phi _{k}^{2}c_{M+Q_{1}+1,k}^{\ast }c_{m+Q_{1}+1,k}  \notag
\\
&&+\sum_{j=0}^{Q_{2}-1}\sum_{k=j+1}^{Q_{2}}\Gamma ^{jk}\left[
c_{N+Q_{1}+1,j}c_{M+Q_{1}+1,j}^{\ast }c_{n+Q_{1}+1,k}^{\ast
}c_{m+Q_{1}+1,k}\right.  \notag \\
&&\left. -\delta _{m,M}\frac{1}{2}c_{N+Q_{1}+1,k}c_{n+Q_{1}+1,k}^{\ast
}-\delta _{n,N}\frac{1}{2}c_{M+Q_{1}+1,k}^{\ast }c_{m+Q_{1}+1,k}\right]
\notag
\end{eqnarray}%
\begin{gather}
W_{B,C}^{\left( N,M,n,m\right) }=\sum_{j,k=0}^{Q_{2}}\Phi _{j}\Phi
_{k}c_{N+Q_{1}+1,j}c_{M+Q_{1}+1,k}^{\ast }c_{n,j}^{\ast
}c_{m+Q_{1}+1,k}-\delta _{m,M}\frac{1}{2}\sum_{k=0}^{Q_{2}}\Phi
_{k}^{2}c_{N+Q_{1}+1,k}c_{n,k}^{\ast }  \notag \\
+\sum_{j=0}^{Q_{2}-1}\sum_{k=j+1}^{Q_{2}}\Gamma ^{jk}\left[
c_{N+Q_{1}+1,j}c_{M+Q_{1}+1,j}^{\ast }c_{n,k}^{\ast }c_{m+Q_{1}+1,k}-\delta
_{m,M}\frac{1}{2}c_{N+Q_{1}+1,k}c_{n,k}^{\ast }\right]
\end{gather}%
\begin{gather}
W_{B,D}^{\left( N,M,n,m\right) }=\sum_{j,k=0}^{Q_{2}}\Phi _{j}\Phi
_{k}c_{N+Q_{1}+1,j}c_{M+Q_{1}+1,k}^{\ast }c_{n+Q_{1}+1,j}^{\ast
}c_{m,k}-\delta _{n,N}\frac{1}{2}\sum_{k=0}^{Q_{2}}\Phi
_{k}^{2}c_{M+Q_{1}+1,k}^{\ast }c_{m,k}  \notag \\
+\sum_{j=0}^{Q_{2}-1}\sum_{k=j+1}^{Q_{2}}\Gamma ^{jk}\left[
c_{N+Q_{1}+1,j}c_{M+Q_{1}+1,j}^{\ast }c_{n+Q_{1}+1,k}^{\ast }c_{m,k}-\delta
_{n,N}\frac{1}{2}c_{M+Q_{1}+1,k}^{\ast }c_{m,k}\right]
\end{gather}%
\begin{gather}
W_{C,A}^{\left( N,M,n,m\right) }=\sum_{j,k=0}^{Q_{2}}\Phi _{j}\Phi
_{k}c_{N,j}c_{M+Q_{1}+1,k}^{\ast }c_{n,j}^{\ast }c_{m,k}-\delta
_{n,N}\sum_{k=0}^{Q_{2}}\frac{1}{2}\Phi _{k}^{2}c_{M+Q_{1}+1,k}^{\ast
}c_{m,k}  \notag \\
+\sum_{j=0}^{Q_{2}-1}\sum_{k=j+1}^{Q_{2}}\Gamma ^{jk}\left[
c_{N,j}c_{M+Q_{1}+1,j}^{\ast }c_{n,k}^{\ast }c_{m,k}-\delta _{n,N}\frac{1}{2}%
c_{M+Q_{1}+1,k}^{\ast }c_{m,k}\right]
\end{gather}%
\begin{gather}
W_{C,B}^{\left( N,M,n,m\right) }=\sum_{j,k=0}^{Q_{2}}\Phi _{j}\Phi
_{k}c_{N,j}c_{M+Q_{1}+1,k}^{\ast }c_{n+Q_{1}+1,j}^{\ast }c_{m+Q_{1}+1,k}
\notag \\
-\delta _{m,M}\sum_{k=0}^{Q_{2}}\frac{1}{2}\Phi
_{k}^{2}c_{N,k}c_{n+Q_{1}+1,k}^{\ast } \\
+\sum_{j=0}^{Q_{2}-1}\sum_{k=j+1}^{Q_{2}}\Gamma ^{jk}\left[
c_{N,j}c_{M+Q_{1}+1,j}^{\ast }c_{n+Q_{1}+1,k}^{\ast }c_{m+Q_{1}+1,k}-\delta
_{M,m}\frac{1}{2}c_{N,k}c_{n+Q_{1}+1,k}^{\ast }\right]  \notag
\end{gather}%
\begin{gather}
W_{C,C}^{\left( N,M,n,m\right) }=\sum_{j,k=0}^{Q_{2}}\Phi _{j}\Phi
_{k}c_{N,j}c_{M+Q_{1}+1,k}^{\ast }c_{n,j}^{\ast }c_{m+Q_{1}+1,k}-\delta
_{m,M}\frac{1}{2}\sum_{k=0}^{Q_{2}}\Phi _{k}^{2}c_{N,k}c_{n,k}^{\ast }
\notag \\
-\delta _{N,n}\sum_{k=0}^{Q_{2}}\frac{1}{2}\Phi
_{k}^{2}c_{M+Q_{1}+1,k}^{\ast }c_{m+Q_{1}+1,k} \\
+\sum_{j=0}^{Q_{2}-1}\sum_{k=j+1}^{Q_{2}}\Gamma ^{jk}\left[
c_{N,j}c_{M+Q_{1}+1,j}^{\ast }c_{n,k}^{\ast }c_{m+Q_{1}+1,k}\right.  \notag
\\
\left. -\delta _{m,M}\frac{1}{2}c_{N,k}c_{n,k}^{\ast }-\delta _{n,N}\frac{1}{%
2}c_{M+Q_{1}+1,k}^{\ast }c_{m+Q_{1}+1,k}\right]  \notag
\end{gather}%
\begin{eqnarray}
W_{C,D}^{\left( N,M,n,m\right) } &=&\sum_{j,k=0}^{Q_{2}}\Phi _{j}\Phi
_{k}c_{N,j}c_{M+Q_{1}+1,k}^{\ast }c_{n+Q_{1}+1,j}^{\ast }c_{m,k} \\
&&+\sum_{j=0}^{Q_{2}-1}\sum_{k=j+1}^{Q_{2}}\Gamma ^{jk}\left[
c_{N,j}c_{M+Q_{1}+1,j}^{\ast }c_{n+Q_{1}+1,k}^{\ast }c_{m,k}\right]  \notag
\end{eqnarray}

When the parameters of the Hamiltonian, and hence the dressed-states, depend
explicitly on time, we make the simplifying assumption that the modulations
and the dissipation are weak, so in evaluating the dressed states one can
use the unperturbed values of the parameters. This is plausible when the
modulation of the parameter $X$ is of the form $X\left( t\right) =X_{0}\left[
1+\varepsilon f\left( t\right) \right] $, where $\left\vert f\left( t\right)
\right\vert \leq 1$ is some time-dependent function and $\varepsilon \ll 1$
is the modulation amplitude. Moreover, to make the system of ODE numerically
treatable, we make an additional approximation whereby we set $%
W_{X,Y}^{\left( N,M,n,m\right) }=0$ whenever $\left\vert W_{X,Y}^{\left(
N,M,n,m\right) }\right\vert <\epsilon $, where $\epsilon $ is the threshold
parameter and $X$ and $Y$ stand for $A$, $B$, $C$ and $D$. This is
reasonable when the dissipation acts as a perturbation to the coherent
dynamics, when all the dissipation rates are much smaller that the
atom--field coupling strength $g$.

\section{Coherent state\label{dodon.coherent}}

For the coherent state $|\alpha \rangle $ with a real amplitude $\alpha $,
we have

\begin{equation}
|\alpha \rangle =e^{-\alpha ^{2}/2}\sum_{n=0}^{\infty }\frac{\alpha ^{2}}{%
\sqrt{n!}}|n\rangle ~.  \label{dodon.cs}
\end{equation}%
The recurrence relations read%
\begin{equation}
\chi _{0}=\exp \left( -\frac{\alpha ^{2}}{2}\right) ~,~\chi _{n+1}=\chi _{n}%
\frac{\alpha }{\sqrt{n+1}}~.
\end{equation}

\begin{figure}[h]
\begin{center}
\includegraphics[width=1.03\textwidth]{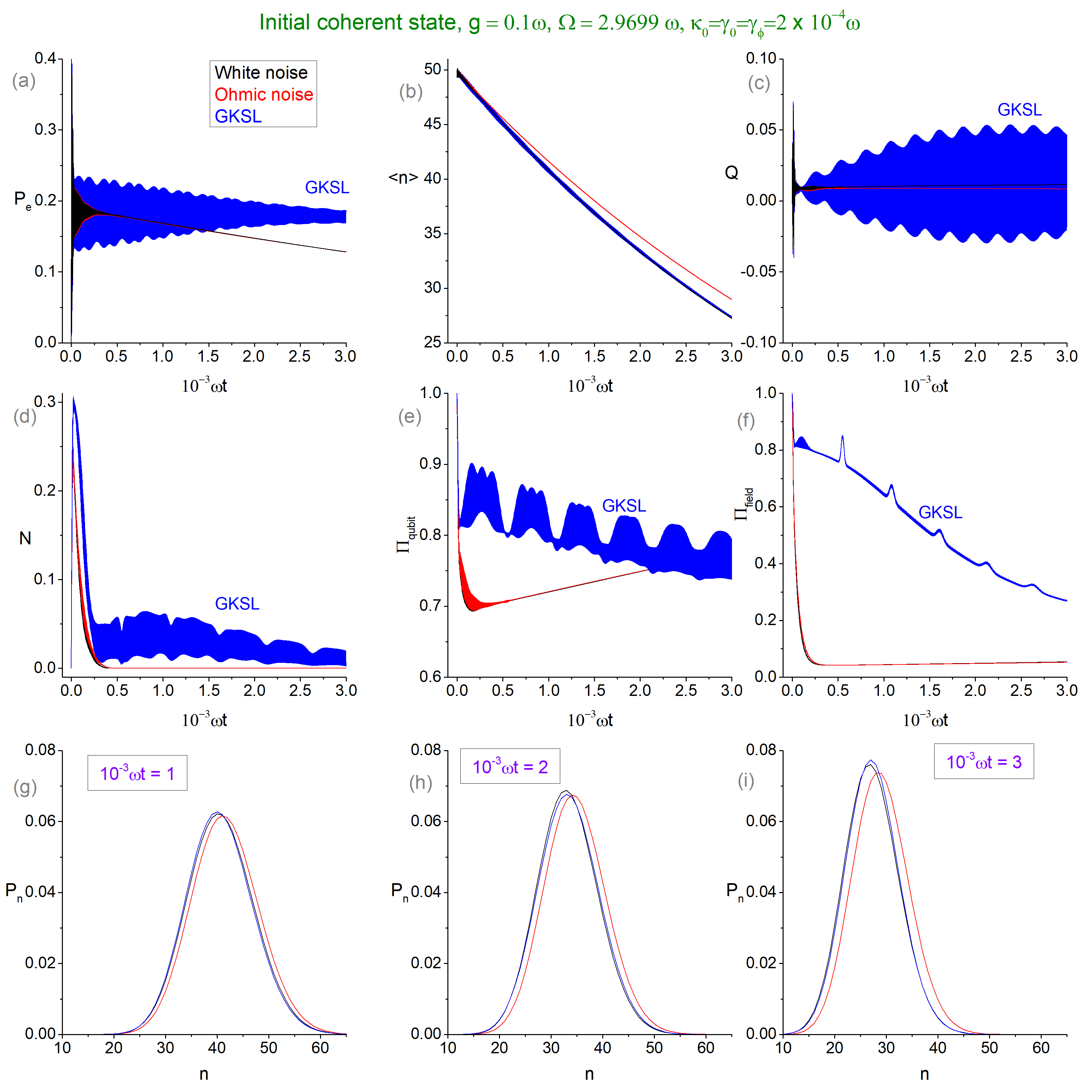} {}
\end{center}
\caption{Behavior of the quantum Rabi model under dissipation for the
initial coherent state, Eq. (\protect\ref{dodon.cs}), and parameters $\protect%
\alpha =\protect\sqrt{50}$, $g=0.1\protect\omega $, $\Omega =2.9699\protect%
\omega $ and $\protect\kappa _{0}=\protect\gamma _{0}=\protect\gamma _{%
\protect\phi }=2\times 10^{-4}\protect\omega $. In all the figures of this
chapter it is assumed that the qubit was initially in the ground state. a)
Qubit excitation probability. b) Average photon number. c) Mandel's $Q$%
-factor. d) Negativity (nonzero values attest qubit-field entanglement). e)
Qubit's purity. f) Field's purity. g) Photon number probability distribution
at the instant of time $\protect\omega t=10^{3}$. h) Photon number
probability distribution at the instant of time $\protect\omega t=2\times
10^{3}$. i) Photon number probability distribution at the instant of time $%
\protect\omega t=3\times 10^{3}$. Blue lines correspond to the predictions
of the GKSL ME, black lines -- to the predictions of the dressed-picture ME
for white noise and red lines -- to the predictions of the dressed-picture
ME for ohmic noise. Whenever the predictions of the GKSL master equation differ significantly from those of the DME, the corresponding curves are explicitly labeled as ``GKSL'' in the plots for improved clarity.}
\label{dodon.fig1}
\end{figure}

\begin{figure}[h]
\begin{center}
\includegraphics[width=1.03\textwidth]{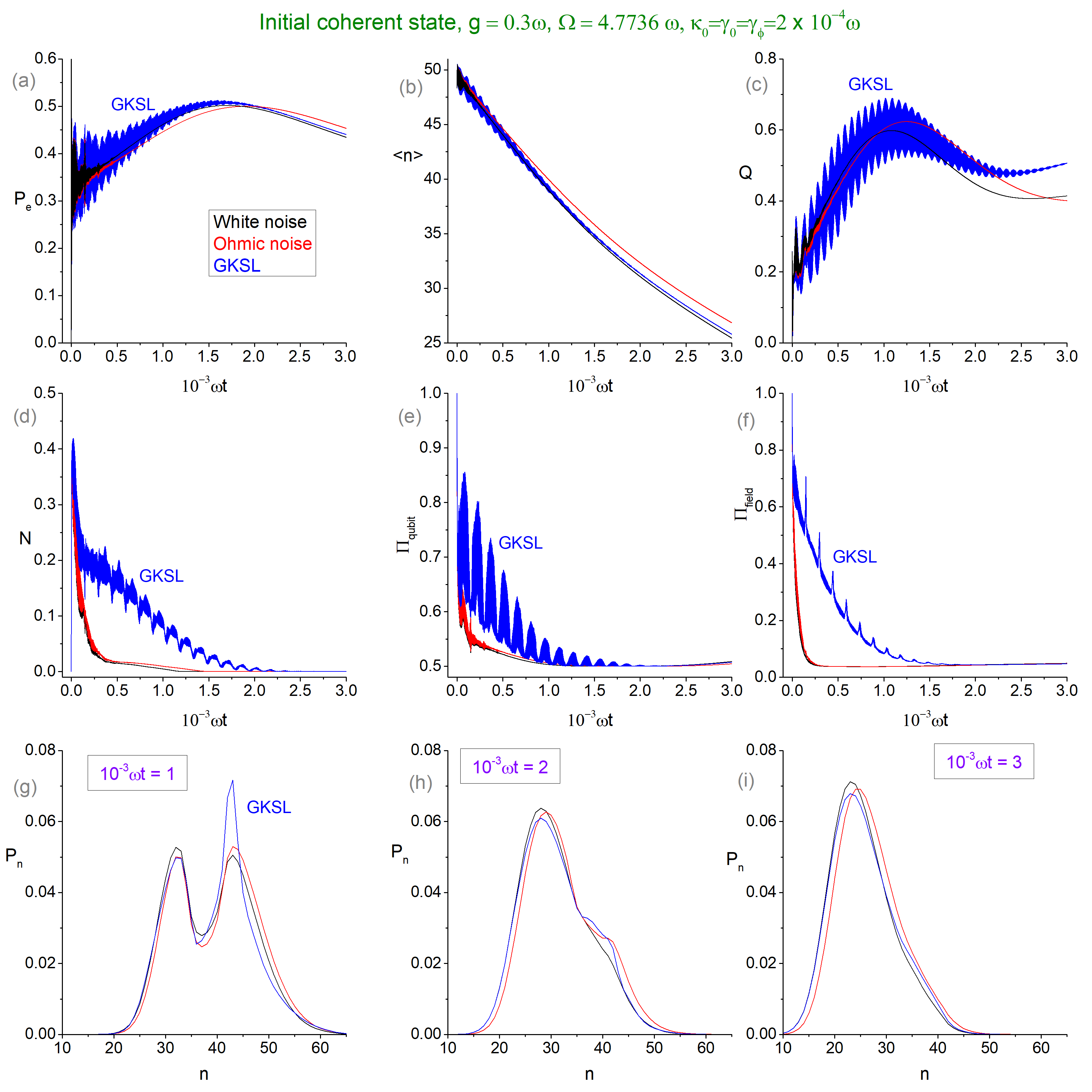} {}
\end{center}
\caption{Similar to Fig. \protect\ref{dodon.fig1} but for parameters $g=0.3%
\protect\omega $ and $\Omega =4.7736\protect\omega $ (in the vicinity of the
$5$-photon resonance).}
\label{dodon.fig2}
\end{figure}

\begin{figure}[h]
\begin{center}
\includegraphics[width=1.03\textwidth]{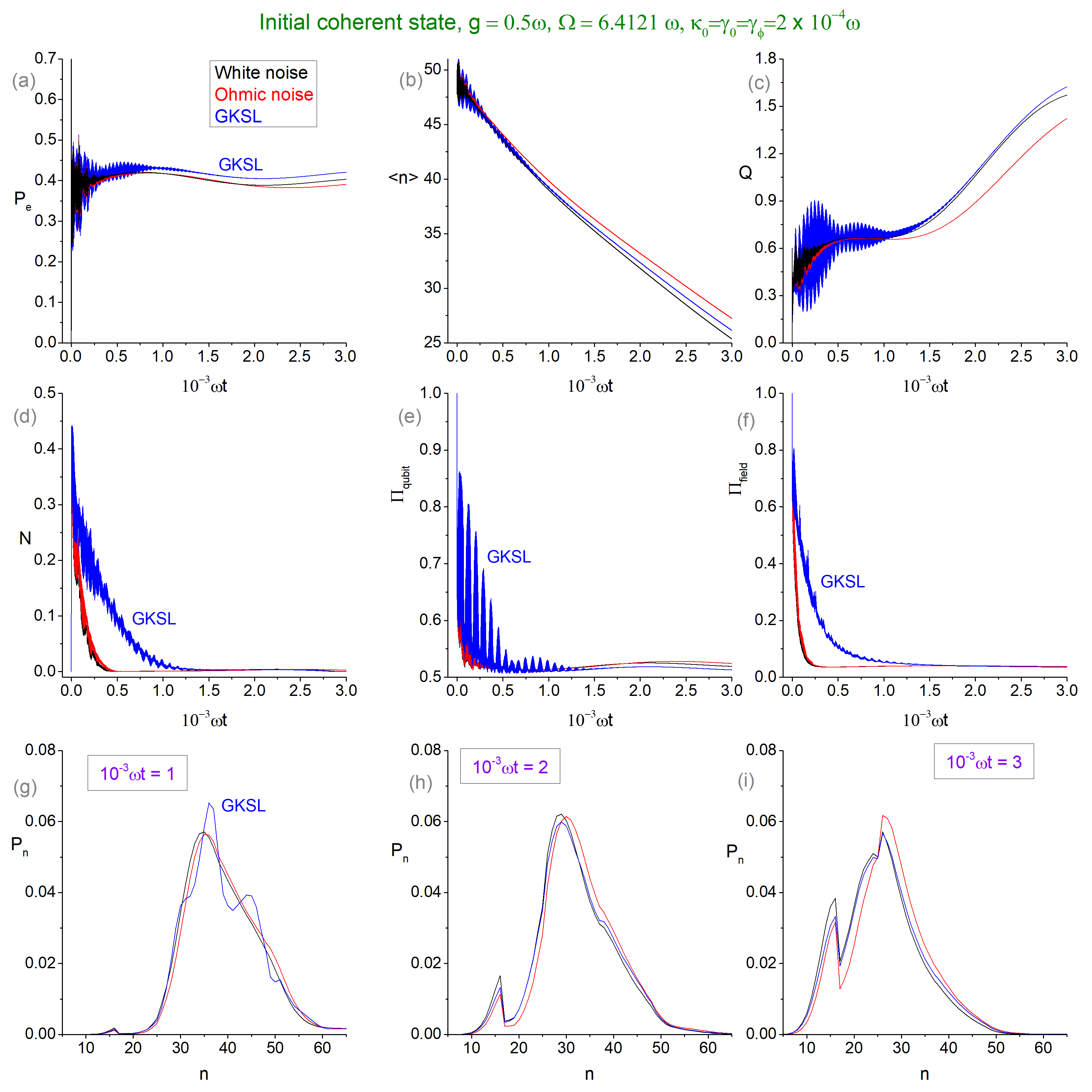} {}
\end{center}
\caption{Similar to Fig. \protect\ref{dodon.fig1} but for parameters $g=0.5%
\protect\omega $ and $\Omega =6.4121\protect\omega $ (in the vicinity of the
$7$-photon resonance).}
\label{dodon.fig3}
\end{figure}

Figures \ref{dodon.fig1}, \ref{dodon.fig2} and \ref{dodon.fig3} show the
dynamics of the atomic excitation probability, $P_{e}=\limfunc{Tr}\left[
\rho |e\rangle \langle e|\right] $, average photon number, $\left\langle
n\right\rangle =\limfunc{Tr}\left[ \rho a^{\dagger }a\right] $, Mandel $Q$%
-factor%
\begin{equation}
Q=\frac{\left\langle \left( \Delta n\right) ^{2}\right\rangle -\left\langle
n\right\rangle }{\left\langle n\right\rangle }~,
\end{equation}%
qubit and cavity purities [Eqs. (\ref{dodon.por1}) and (\ref{dodon.por2})], negativity [Eq. \ref{dodon.nee}], and the photon number probability
distribution for three different time instants. The parameters were chosen
to correspond, approximately, to the three-, five- and seven-photon
resonances, as will be explained in Section \ref{dodon.multi} (note that the exact
resonances are photon number dependent \cite%
{dodon.multi1,dodon.multi2,dodon.multi3,dodon.multi4}).

First of all, we can observe that the GKSL ME (blue lines) predicts much
faster damping of the fast oscillations (due to the exchange of excitations
between the qubit and the field), which on the considered time interval are
perceived as a large width of the curves. Besides, DME for both ohmic and
white spectral densities predict much faster decay of the purities and
negativity.This observation seems to agree with the conclusion of \cite%
{dodon.Settineri} that in the dispersive regime DME overestimates the
effects of relaxation. Moreover, the predictions of the three models for $%
P_{e}$, $\left\langle n\right\rangle $ and $Q$ do not coincide for large
times, $gt\gg 1$, meaning that the accurate description of the system
requires a precise knowledge of the spectral density in order to use an
appropriate master equation.

\section{Odd Schr\"{o}dinger cat state\label{dodon.cat}}

The odd Schr\"{o}dinger cat state \cite{dodon.knight} is defined in terms of
the coherent states with opposite amplitudes as
\begin{equation}
|\psi \rangle =\frac{|\alpha \rangle -|-\alpha \rangle }{\sqrt{%
2-2e^{-2\alpha ^{2}}}}~,  \label{dodon.oss}
\end{equation}%
where for brevity we consider that $\alpha $ is a real parameter. The
recurrence relations are:%
\begin{equation}
\chi _{0}=0~,~\chi _{1}=\sqrt{\frac{2}{1-e^{-2\alpha ^{2}}}}\alpha
e^{-\alpha ^{2}/2}~,~\chi _{n+2}=\chi _{n}\frac{\alpha ^{2}}{\sqrt{\left(
n+2\right) \left( n+1\right) }}\,.
\end{equation}

\begin{figure}[h]
\begin{center}
\includegraphics[width=1.03\textwidth]{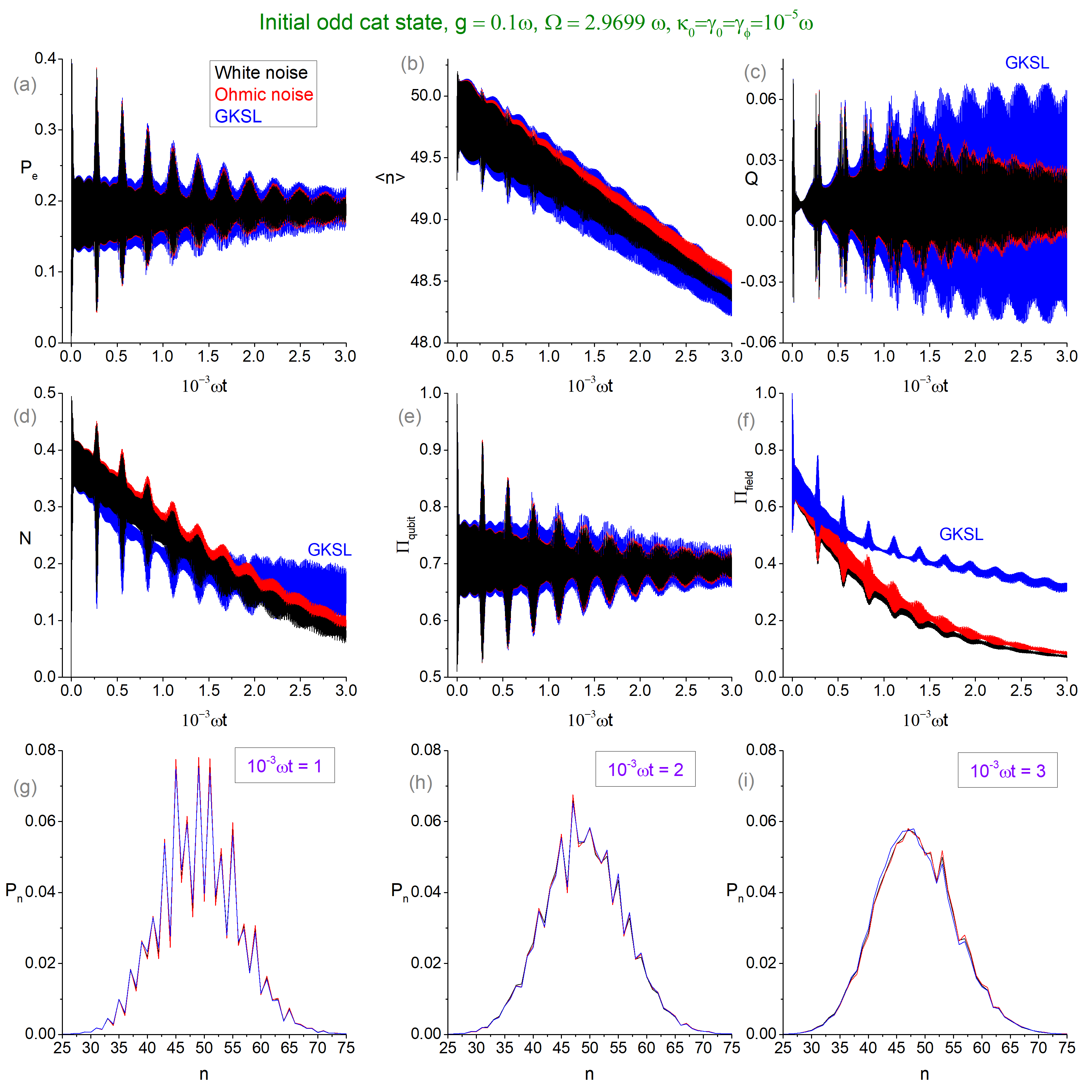} {}
\end{center}
\caption{Behavior of the quantum Rabi model under dissipation for the
initial odd Schr\"{o}dinger cat state, Eq. (\protect\ref{dodon.oss}), and
parameters $\protect\alpha =7.071$ (corresponding to $\langle n(0)\rangle
=50 $), $g=0.1\protect\omega $, $\Omega =2.9699\protect\omega $ and $\protect%
\kappa _{0}=\protect\gamma _{0}=\protect\gamma _{\protect\phi }=10^{-5}%
\protect\omega $.}
\label{dodon.fig4}
\end{figure}

The results for the Schr\"{o}dinger cat states with the initial average
photon number $\left\langle n\left( 0\right) \right\rangle =50$
(corresponding to $\alpha =7.071$) and coupling constants $g=0.1\omega $, $%
0.3\omega $ and $0.5\omega $ are shown in Figures \ref{dodon.fig4} -- \ref%
{dodon.fig6}. We see that the predictions of the different dissipative
models for the quantities $P_{e}$, $\left\langle n\right\rangle $ and $Q$
are similar (although DME predicts faster relaxation of the coherences). On
the other hand, for the quantities $\mathcal{N}$, $\Pi _{qubit}$ and $\Pi
_{field}$, as well as the photon number statistics, the differences may be
substantial, depending on the parameters and the time interval.

\begin{figure}[h]
\begin{center}
\includegraphics[width=1.03\textwidth]{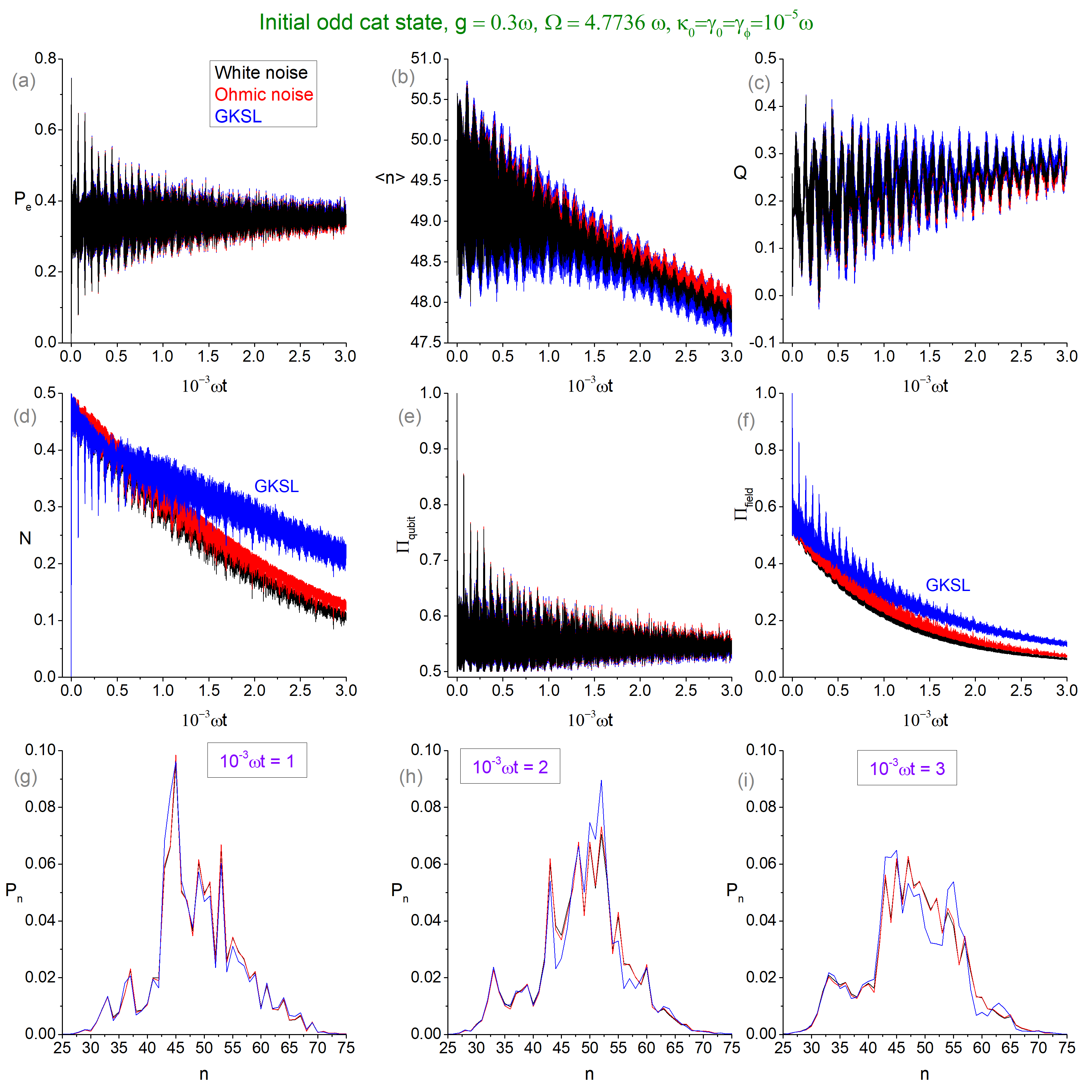} {}
\end{center}
\caption{Similar to Fig. \protect\ref{dodon.fig4} but for parameters $g=0.3%
\protect\omega $ and $\Omega =4.7736\protect\omega $.}
\label{dodon.fig5}
\end{figure}

\begin{figure}[h]
\begin{center}
\includegraphics[width=1.03\textwidth]{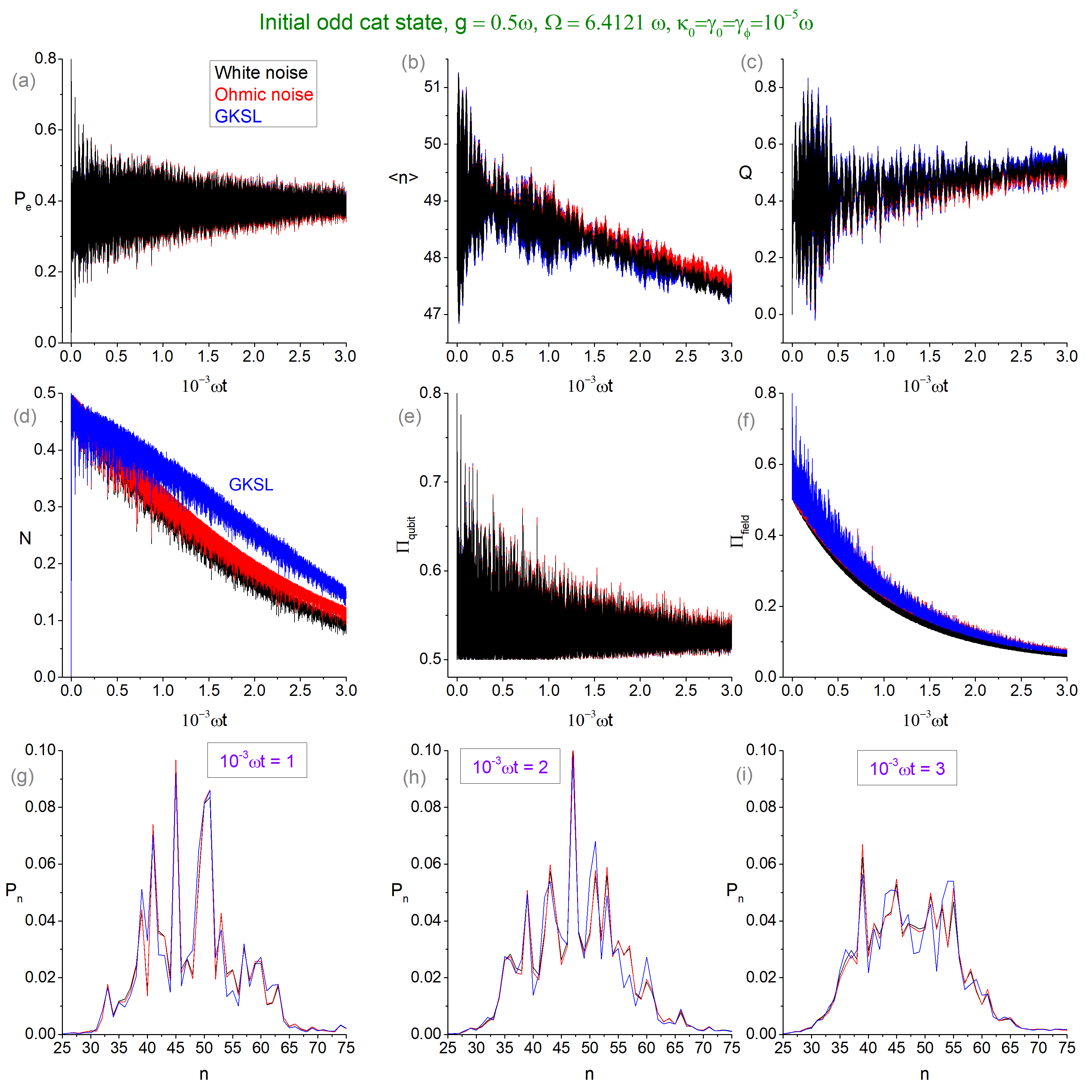} {}
\end{center}
\caption{Similar to Fig. \protect\ref{dodon.fig4} but for parameters $g=0.5%
\protect\omega $ and $\Omega =6.4121\protect\omega $.}
\label{dodon.fig6}
\end{figure}

\section{Squeezed coherent state\label{dodon.squeezedc}}

For the real parameters $s$ and $\beta $, the probability amplitude of the
squeezed coherent state $|s,\beta \rangle $ is%
\begin{equation}
\chi _{n}=\exp \left[ -\frac{\beta ^{2}}{2}\left( 1-\tanh s\right) \right]
\frac{1}{\sqrt{n!\cosh s}}\left( \frac{\tanh s}{2}\right) ^{n/2}H_{n}\left(
x\right) ~,
\end{equation}%
where $H_{n}\left( x\right) $ is the Hermite polynomial and $x=\beta /\sqrt{%
\sinh 2s}$ \cite{dodon.knight,dodon.scully,dodon.vogel,dodon.orszag}. The
recurrence relations are%
\begin{equation}
\chi _{0}=\exp \left[ -\frac{1}{2}\beta ^{2}\left( 1-\tanh s\right) \right]
\frac{1}{\sqrt{\cosh s}}~,~\chi _{n+1}=\chi _{n}K_{n}\left( x\right) \sqrt{%
\frac{\tanh s}{2\left( n+1\right) }}~.  \label{dodon.rr}
\end{equation}%
Here we defined $K_{n}\left( x\right) \equiv H_{n+1}\left( x\right)
/H_{n}\left( x\right) $, and from the well known recurrence relations for
the Hermite polynomials, one has%
\begin{equation}
K_{0}\left( x\right) =2x~,~K_{n+1}\left( x\right) =2x-\frac{2\left(
n+1\right) }{K_{n}\left( x\right) }~.
\end{equation}

\begin{figure}[h]
\begin{center}
\includegraphics[width=1.03\textwidth]{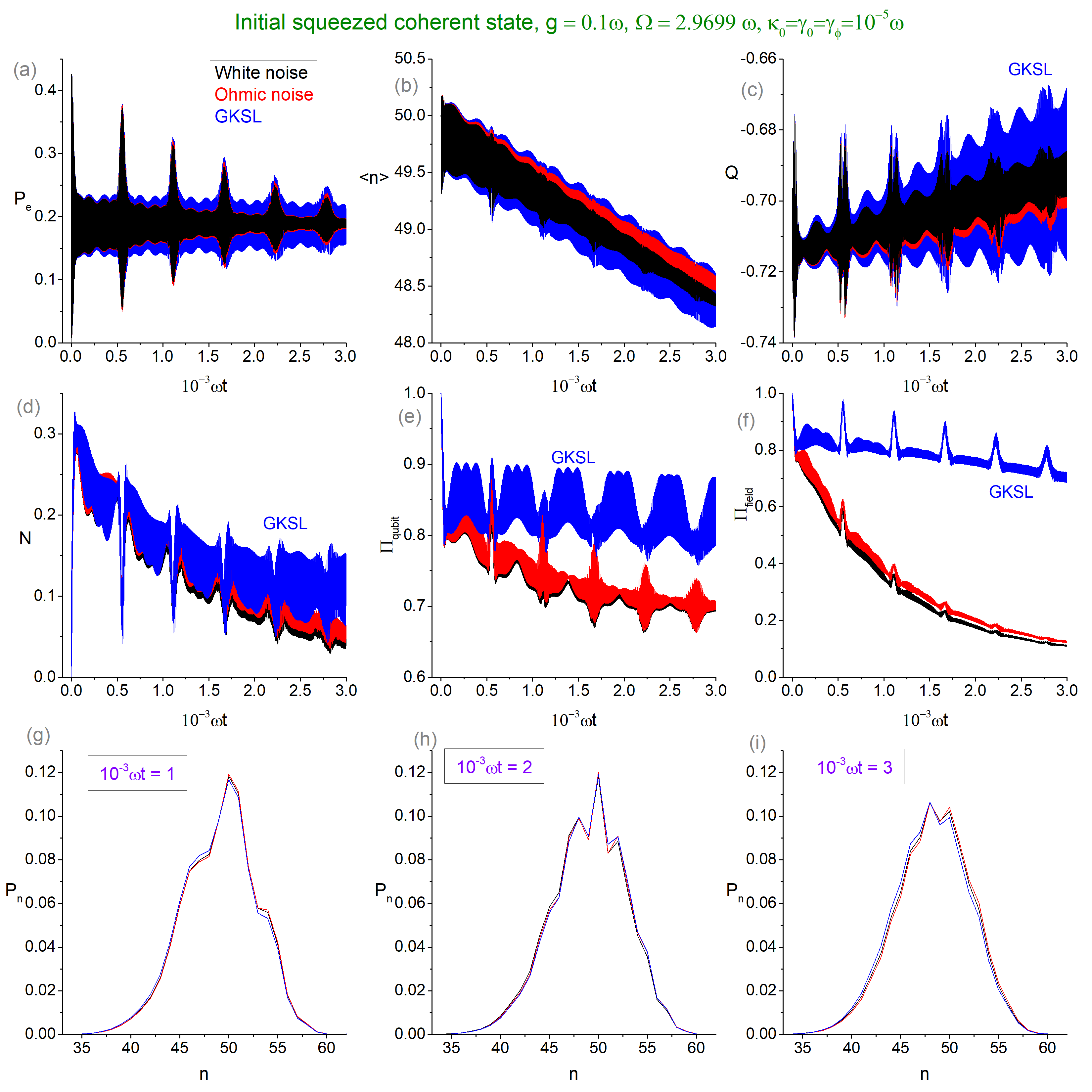} {}
\end{center}
\caption{Behavior of the quantum Rabi model under dissipation for the
initial squeezed coherent state, Eq. (\protect\ref{dodon.rr}), and parameters $%
s=0.7$, $\protect\beta =14.157$ (corresponding to $\langle n(0)\rangle =50$%
), $g=0.1\protect\omega $, $\Omega =2.9699\protect\omega $ and $\protect%
\kappa _{0}=\protect\gamma _{0}=\protect\gamma _{\protect\phi }=10^{-5}%
\protect\omega $.}
\label{dodon.fig7}
\end{figure}

\begin{figure}[h]
\begin{center}
\includegraphics[width=1.03\textwidth]{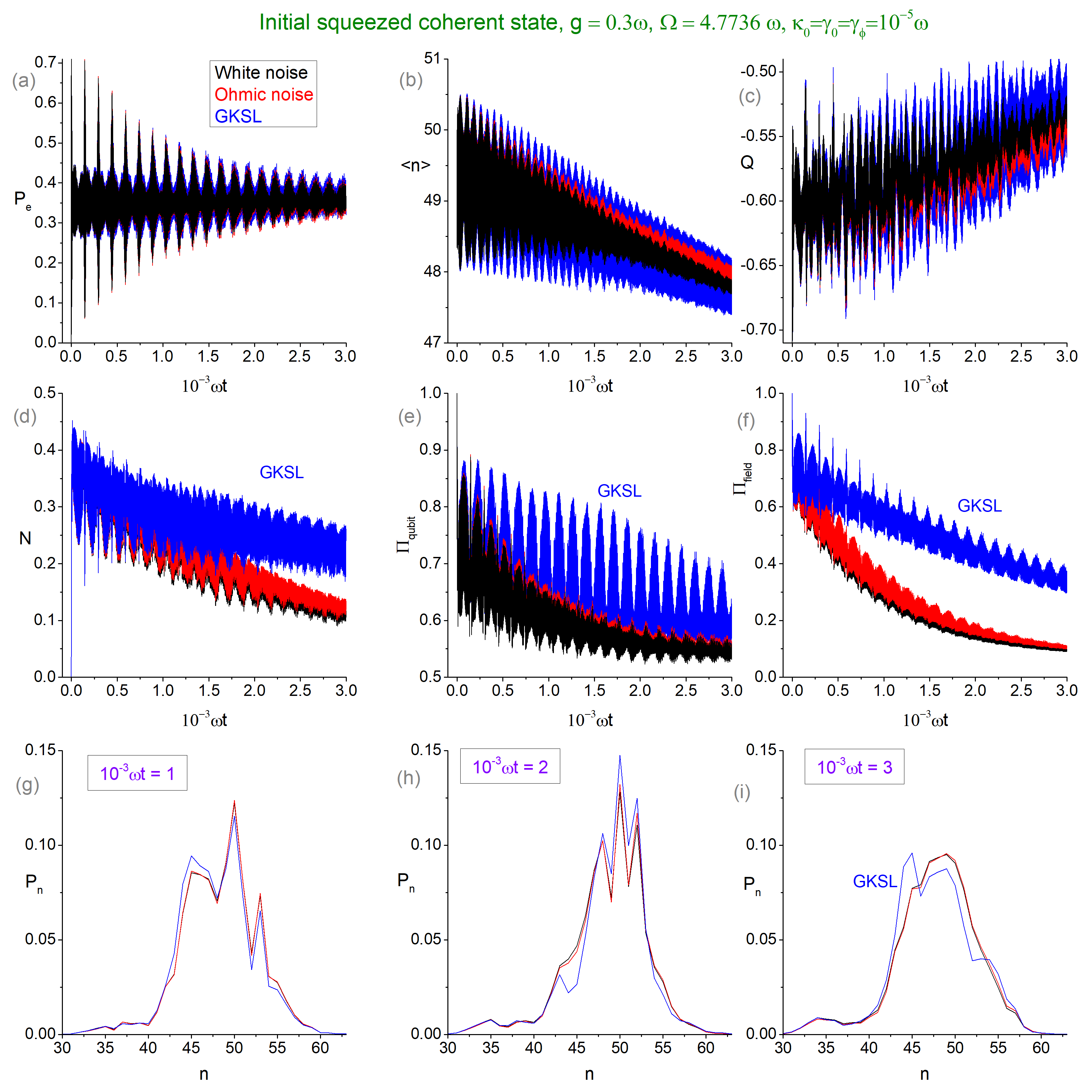} {}
\end{center}
\caption{Similar to Fig. \protect\ref{dodon.fig7} but for parameters $g=0.3%
\protect\omega $ and $\Omega =4.7736\protect\omega $.}
\label{dodon.fig8}
\end{figure}

\begin{figure}[h]
\begin{center}
\includegraphics[width=1.03\textwidth]{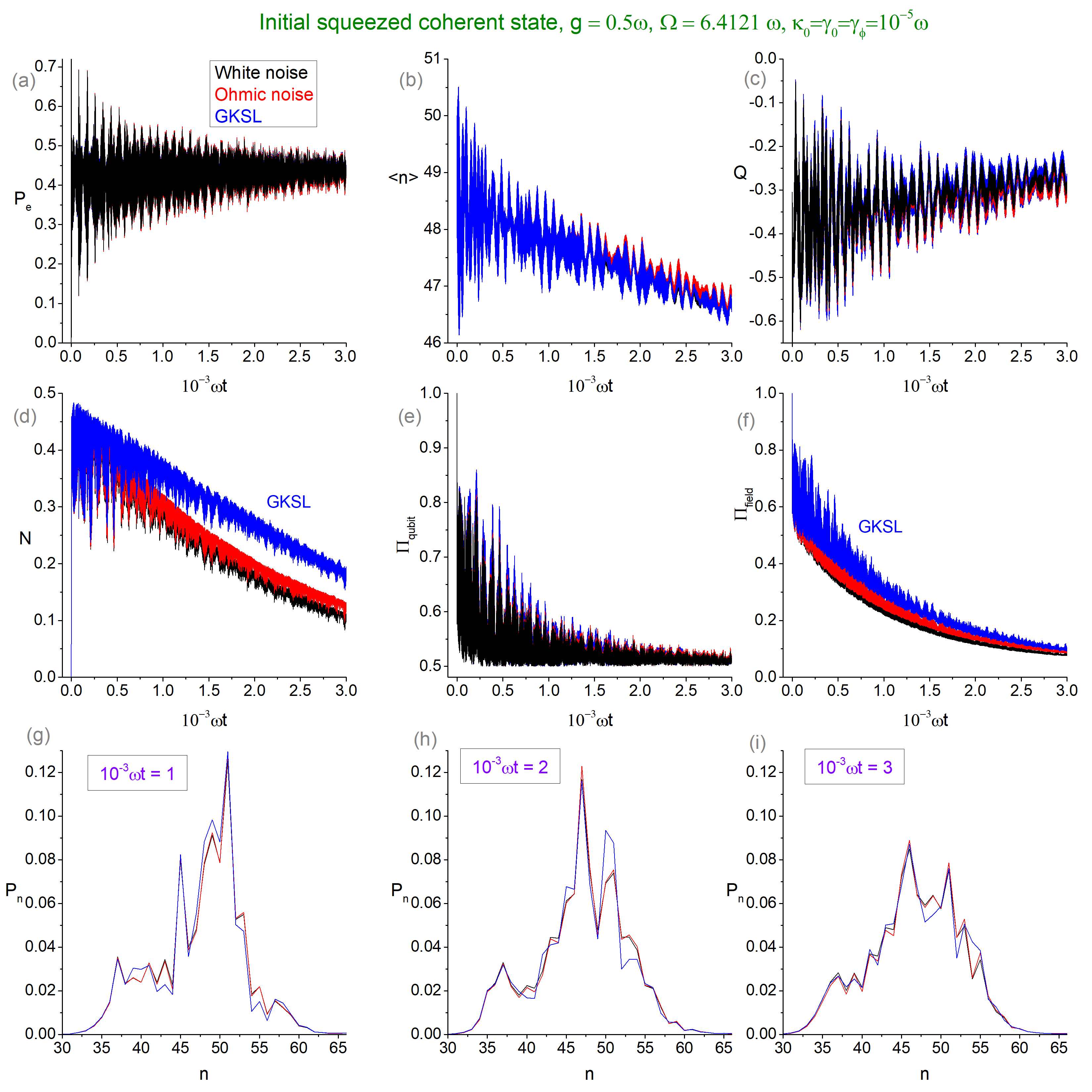} {}
\end{center}
\caption{Similar to Fig. \protect\ref{dodon.fig7} but for parameters $g=0.5%
\protect\omega $ and $\Omega =6.4121\protect\omega $.}
\label{dodon.fig9}
\end{figure}

The results for the initial average photon number $\left\langle n\left(
0\right) \right\rangle =50$ (corresponding to $s=0.7$ and $\beta =14.157$)
and coupling constants $g=0.1\omega $, $0.3\omega $ and $0.5\omega $ are
shown in Figures \ref{dodon.fig7} -- \ref{dodon.fig9}. From the plots we
obtain, similar conclusions to the ones in Section \ref{dodon.cat}.

\section{Squeezed vacuum state\label{dodon.squeezedv}}

If $\beta =0$, the recurrence relations (\ref{dodon.rr}) become%
\begin{equation}
\chi _{0}=\frac{1}{\sqrt{\cosh s}}~,~\chi _{1}=0~,~\chi _{m+2}=-\chi _{m}%
\sqrt{\frac{m+1}{m+2}}\tanh s~.  \label{dodon.svs}
\end{equation}

\begin{figure}[h]
\begin{center}
\includegraphics[width=1.03\textwidth]{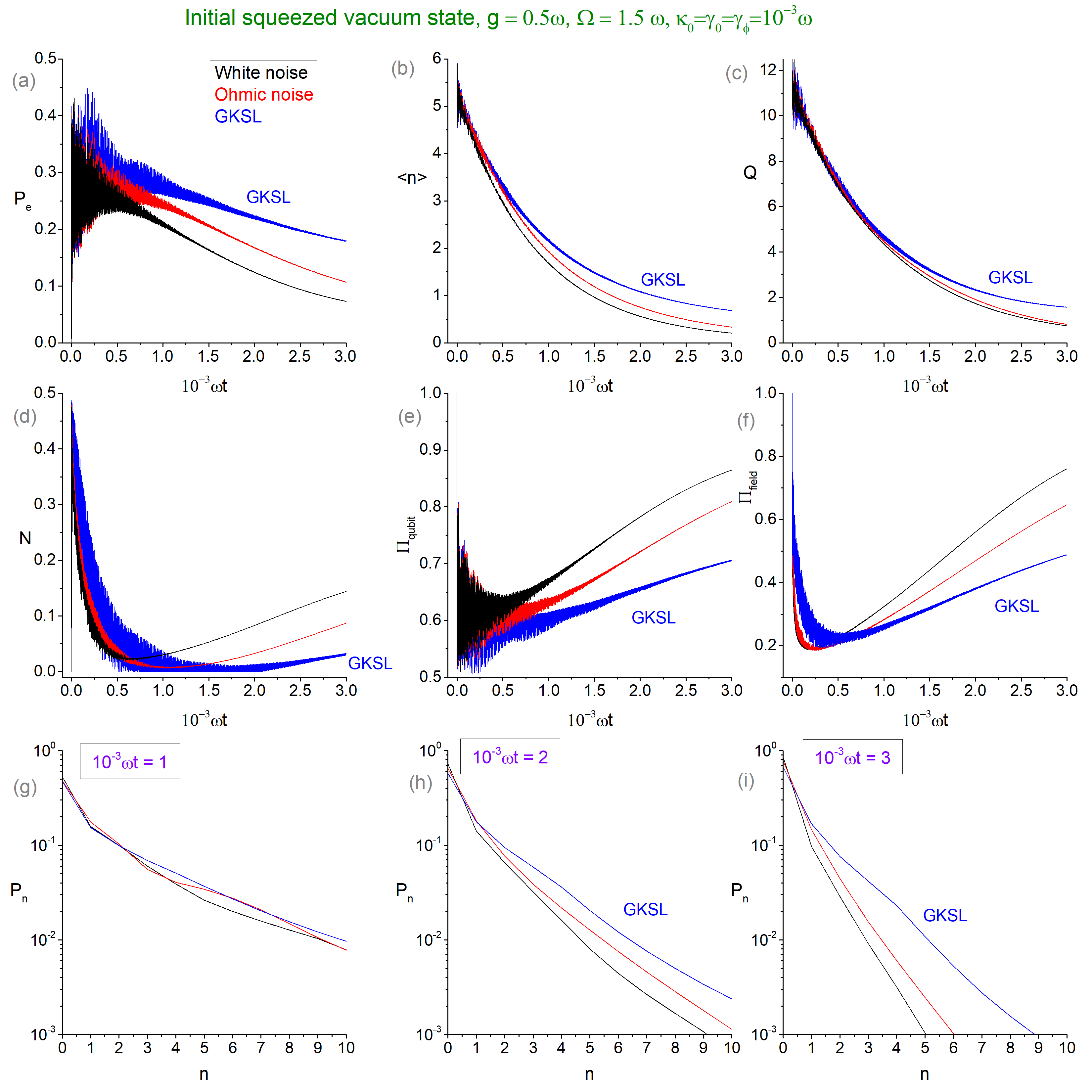} {}
\end{center}
\caption{Behavior of the quantum Rabi model under dissipation for the
initial squeezed vacuum state, Eq. (\protect\ref{dodon.svs}), and parameters $%
s=1.544$ (corresponding to $\langle n(0)\rangle =5$), $g=0.5\protect\omega $%
, $\Omega =1.5\protect\omega $ (near resonant regime) and $\protect\kappa %
_{0}=\protect\gamma _{0}=\protect\gamma _{\protect\phi }=10^{-3}\protect%
\omega $.}
\label{dodon.fig10}
\end{figure}

\begin{figure}[h]
\begin{center}
\includegraphics[width=1.03\textwidth]{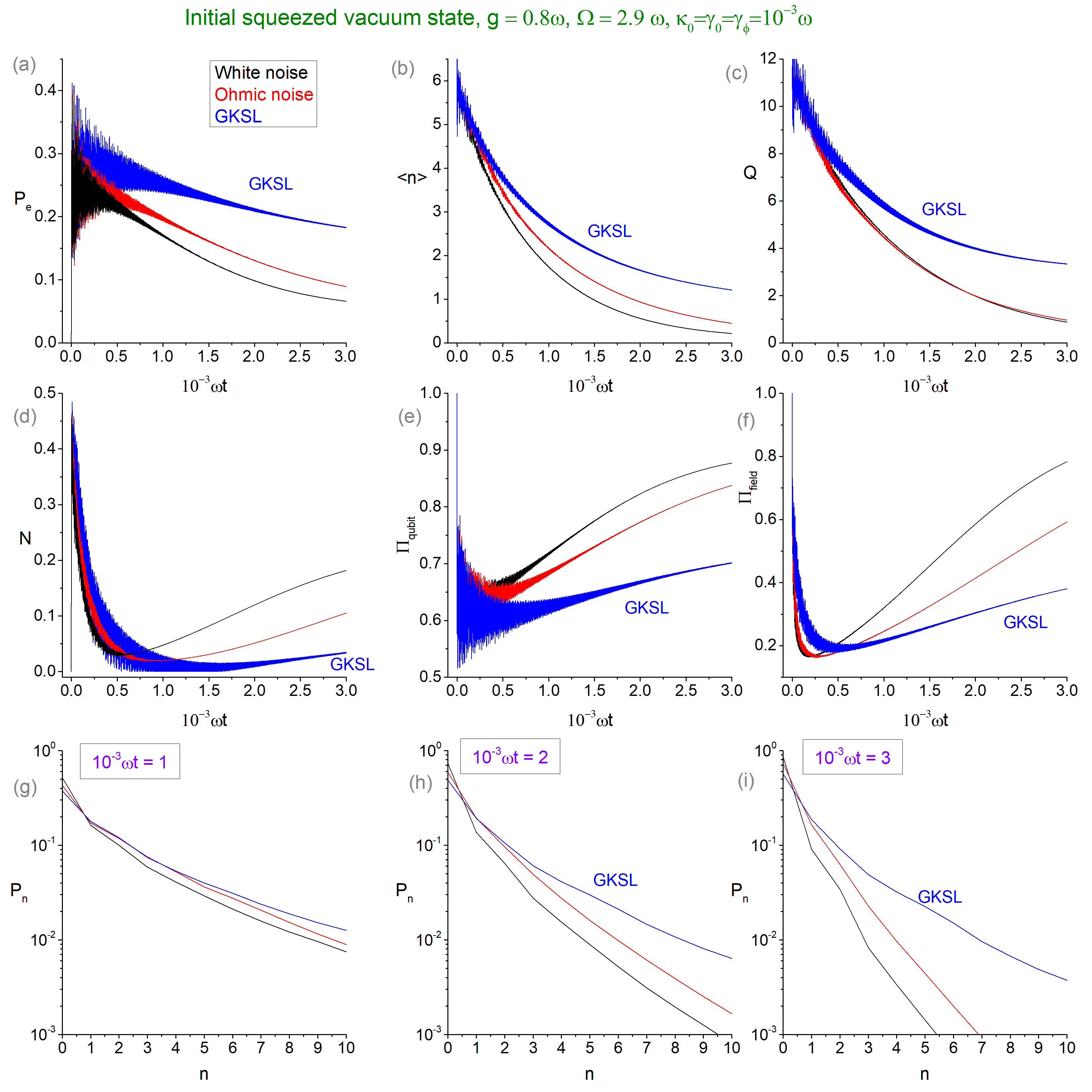} {}
\end{center}
\caption{Similar to Fig. \protect\ref{dodon.fig10} but for parameters $g=0.8%
\protect\omega $ and $\Omega =2.9\protect\omega $.}
\label{dodon.fig11}
\end{figure}

We show the results for the initial state with $\left\langle n\left(
0\right) \right\rangle =5$ (corresponding to the squeezing parameter $%
s=1.544 $) in Figures \ref{dodon.fig10} (for $g=0.5\omega $, $\Omega
=1.5\omega $) and \ref{dodon.fig11} (for $g=0.8\omega $, $\Omega =2.9\omega $%
), assuming relatively strong dissipative rates $\kappa _{0}=\gamma
_{0}=\gamma _{ph}=10^{-3}\omega $. In all these examples, the discrepancies
between the three dissipative models are substantial, therefore an accurate
description of the system dynamics require a precise knowledge of the
spectral density of the reservoir.

\section{Thermal state\label{dodon.thermal}}

For the thermal state, which is a mixed state, the cavity density operator is%
\begin{equation}
\rho _{c}\left( 0\right) =\sum_{n=0}^{\infty }\chi _{n}|n\rangle \langle
n|~,~\chi _{n}=\frac{\alpha ^{n}}{\left( \alpha +1\right) ^{n+1}}~,
\label{dodon.th}
\end{equation}%
where $\alpha $ is the average number of photons. The recurrence relations
are%
\begin{equation}
\chi _{0}=\frac{1}{\alpha +1}~,~\chi _{n+1}=\chi _{n}\frac{\alpha }{\alpha +1%
}~,
\end{equation}%
while the initial condition (\ref{dodon.beh}) becomes%
\begin{equation}
r_{n,m}\left( 0\right) =\sum_{i=0}^{Q_{1}}\chi _{i}c_{i+1,n}^{\ast }c_{i+1,m}\,.
\end{equation}

The system behavior for $\alpha =5$ and other parameters as in Figures \ref%
{dodon.fig10} -- \ref{dodon.fig11} is illustrated in Figures \ref%
{dodon.fig12} -- \ref{dodon.fig13}. As in Section \ref{dodon.squeezedv}, the
discrepancies between the predictions of the dissipative models are
significant for all the considered quantities.

\begin{figure}[!h]
\begin{center}
\includegraphics[width=1.03\textwidth]{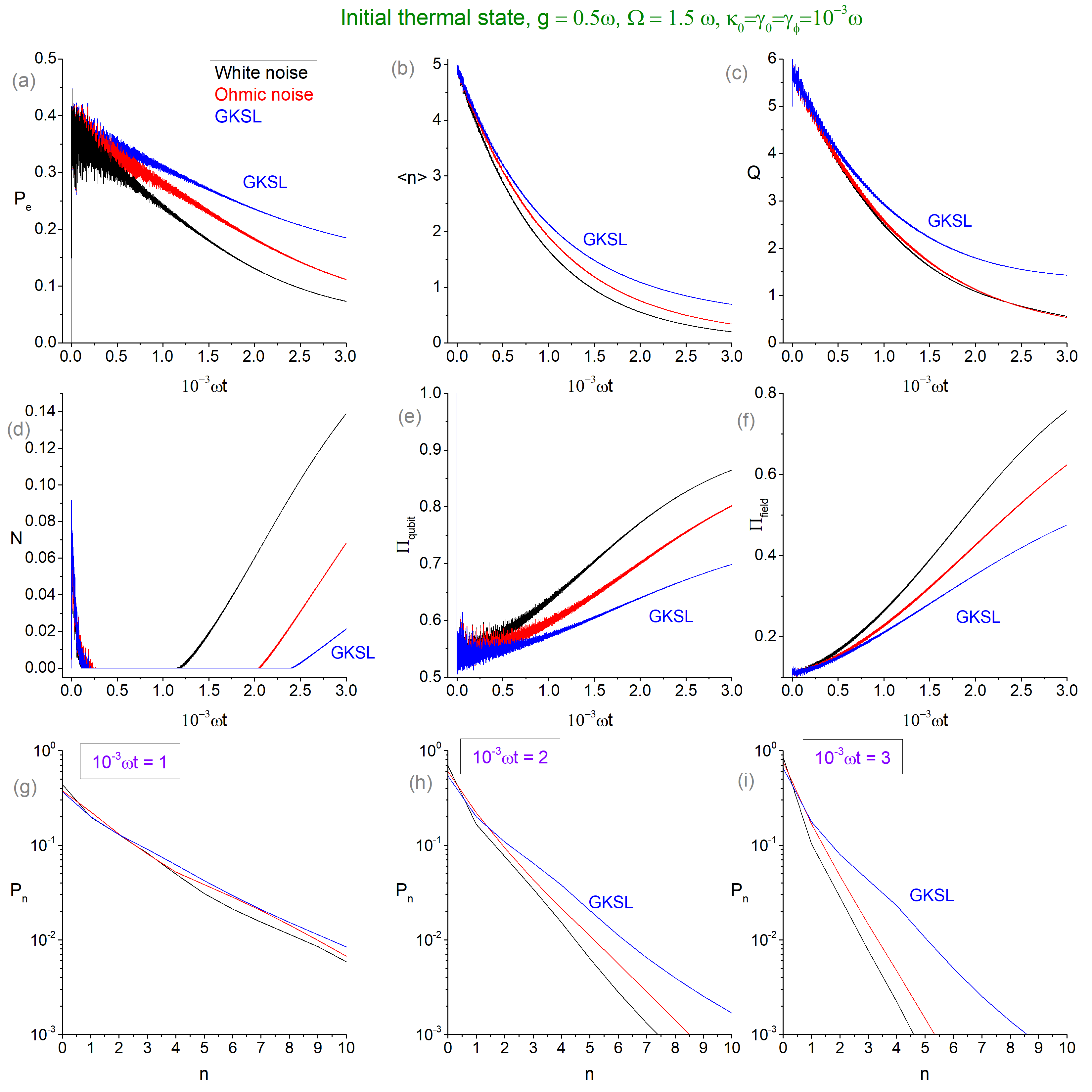} {}
\end{center}
\caption{Behavior of the quantum Rabi model under dissipation for the
initial thermal state, Eq. (\protect\ref{dodon.th}), and parameters $\protect\alpha%
=5$, $g=0.5\protect\omega$, $\Omega=1.5\protect\omega$ and $\protect\kappa_0=%
\protect\gamma_0=\protect\gamma_\protect\phi=10^{-3}\protect\omega$.}
\label{dodon.fig12}
\end{figure}

\begin{figure}[h]
\begin{center}
\includegraphics[width=1.03\textwidth]{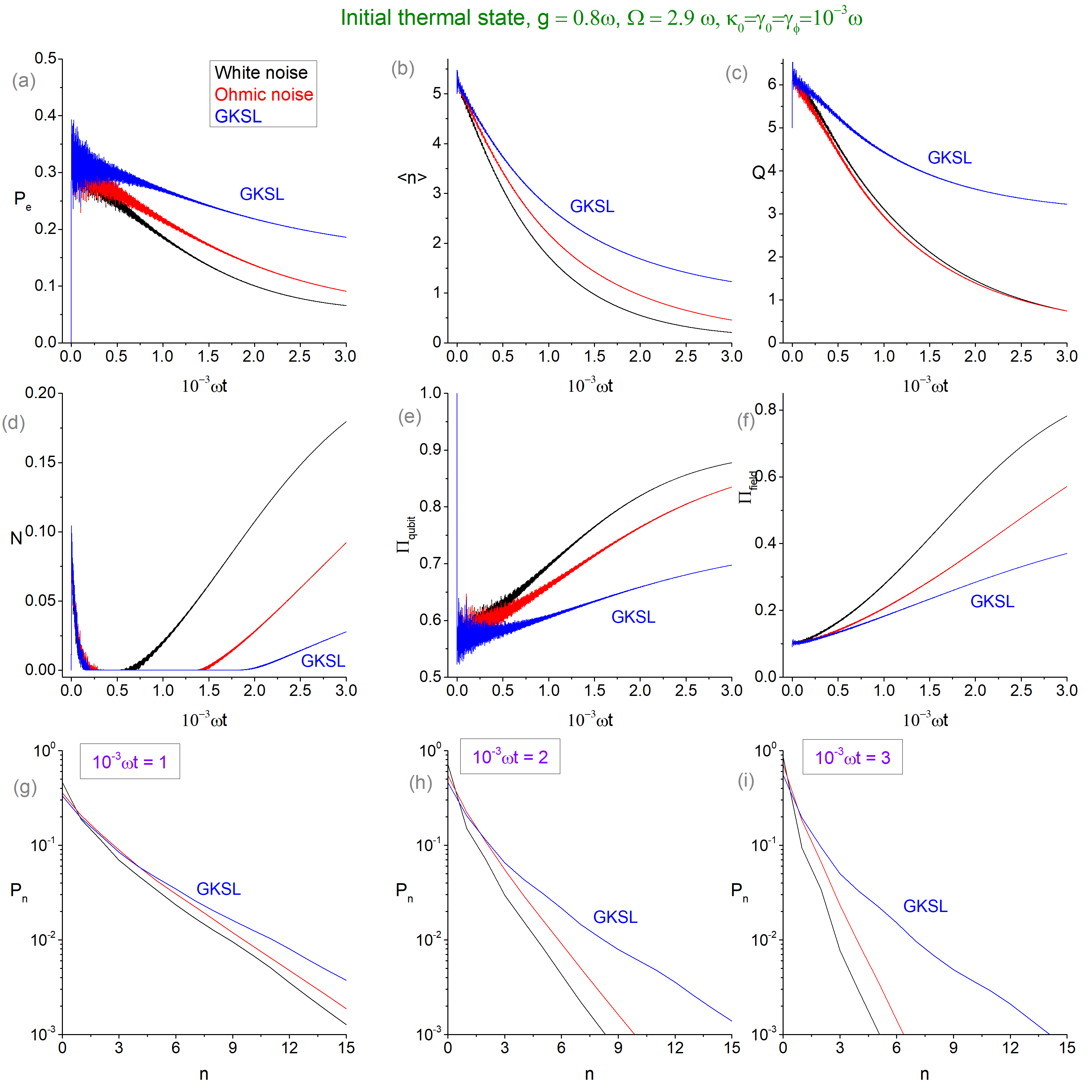} {}
\end{center}
\caption{Similar to Fig. \protect\ref{dodon.fig12} but for parameters $g=0.8%
\protect\omega $ and $\Omega =2.9\protect\omega $.}
\label{dodon.fig13}
\end{figure}

\section{Multi-photon Rabi oscillations\label{dodon.multi}}

\begin{figure}[h]
\begin{center}
\includegraphics[width=1.03\textwidth]{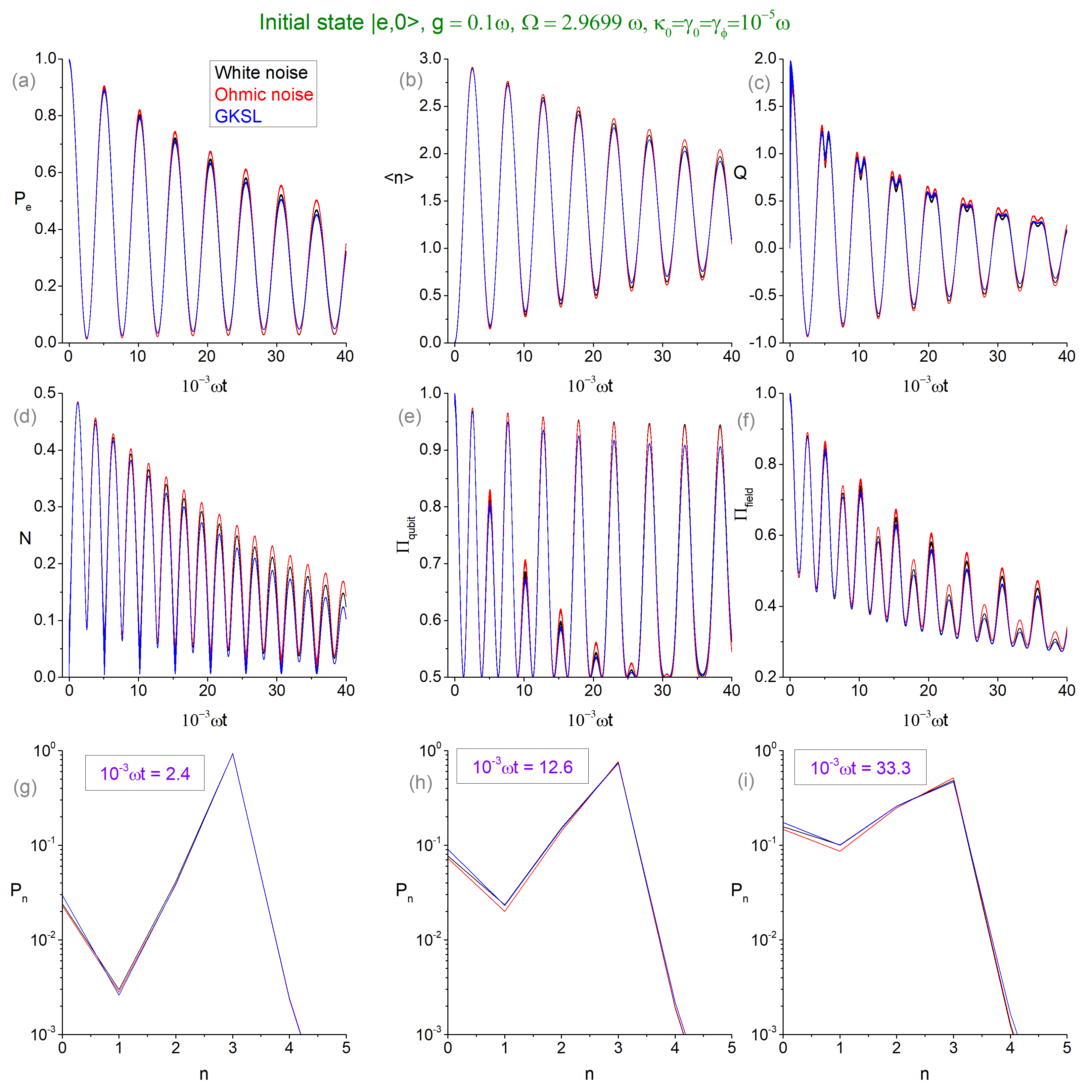} {}
\end{center}
\caption{Similar to Fig. \protect\ref{dodon.fig13} but for the initial state
$|e,0\rangle $ and the parameters corresponding to the three-photon
light--matter resonance: $g=0.1\protect\omega $ and $\Omega =2.9699\protect%
\omega $.}
\label{dodon.fig14}
\end{figure}

Now we consider the initial state $|e,0\rangle $ and study the deexcitation
of the qubit via 3-, 5- and 7-photon processes, which are intrinsic to both
the quantum and semiclassical Rabi Hamiltonian due to the counter-rotating
terms \cite%
{dodon.multi1,dodon.multi2,dodon.multi3,dodon.multi4,dodon.klim0,dodon.3fot1,dodon.3fot2,dodon.3fot3}%
. In Figure \ref{dodon.fig14} we consider the three-photon resonance
(betweem the states $|e,0\rangle $ and $|g,3\rangle $) for the parameters $%
g=0.1\omega $, $\Omega =2.9699\omega $ and the dissipative rates $\kappa
_{0}=\gamma _{0}=\gamma _{ph}=10^{-5}\omega $. By plotting the same
quantities as in the previous figures, we see that the predictions of the
three dissipative models are very close, although the DME predicts a
slightly faster decay of the coherences (as in the previous cases).
Moreover, panels \ref{dodon.fig14}g -- \ref{dodon.fig14}i confirm that the
deexcitation of the qubit populates the cavity mostly in the 3-photon Fock
state.

\begin{figure}[h]
\begin{center}
\includegraphics[width=1.03\textwidth]{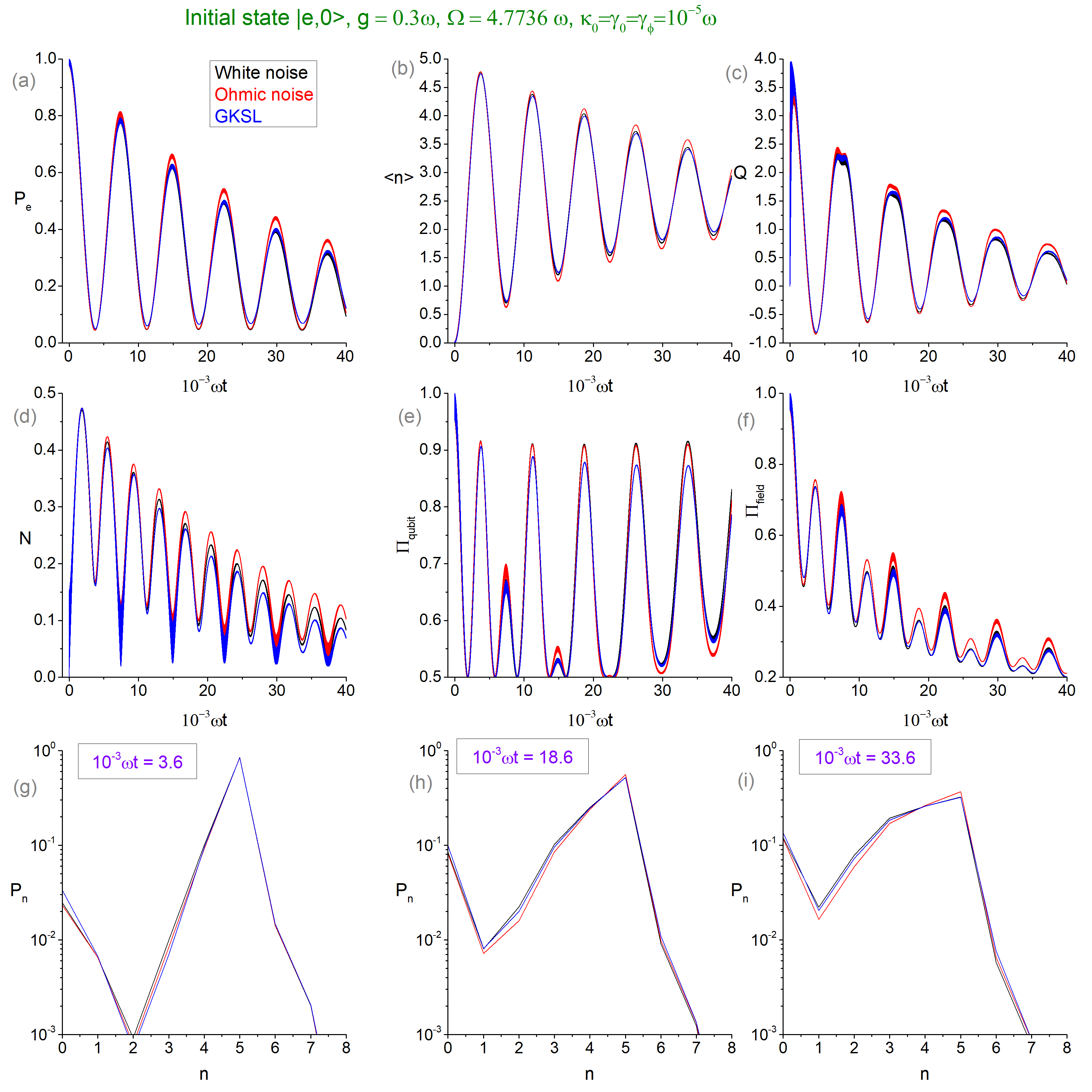} {}
\end{center}
\caption{Similar to Fig. \protect\ref{dodon.fig14}, but for the five-photon
light--matter resonance: $g=0.3\protect\omega $ and $\Omega =4.7736\protect%
\omega $.}
\label{dodon.fig15}
\end{figure}

\begin{figure}[h]
\begin{center}
\includegraphics[width=1.03\textwidth]{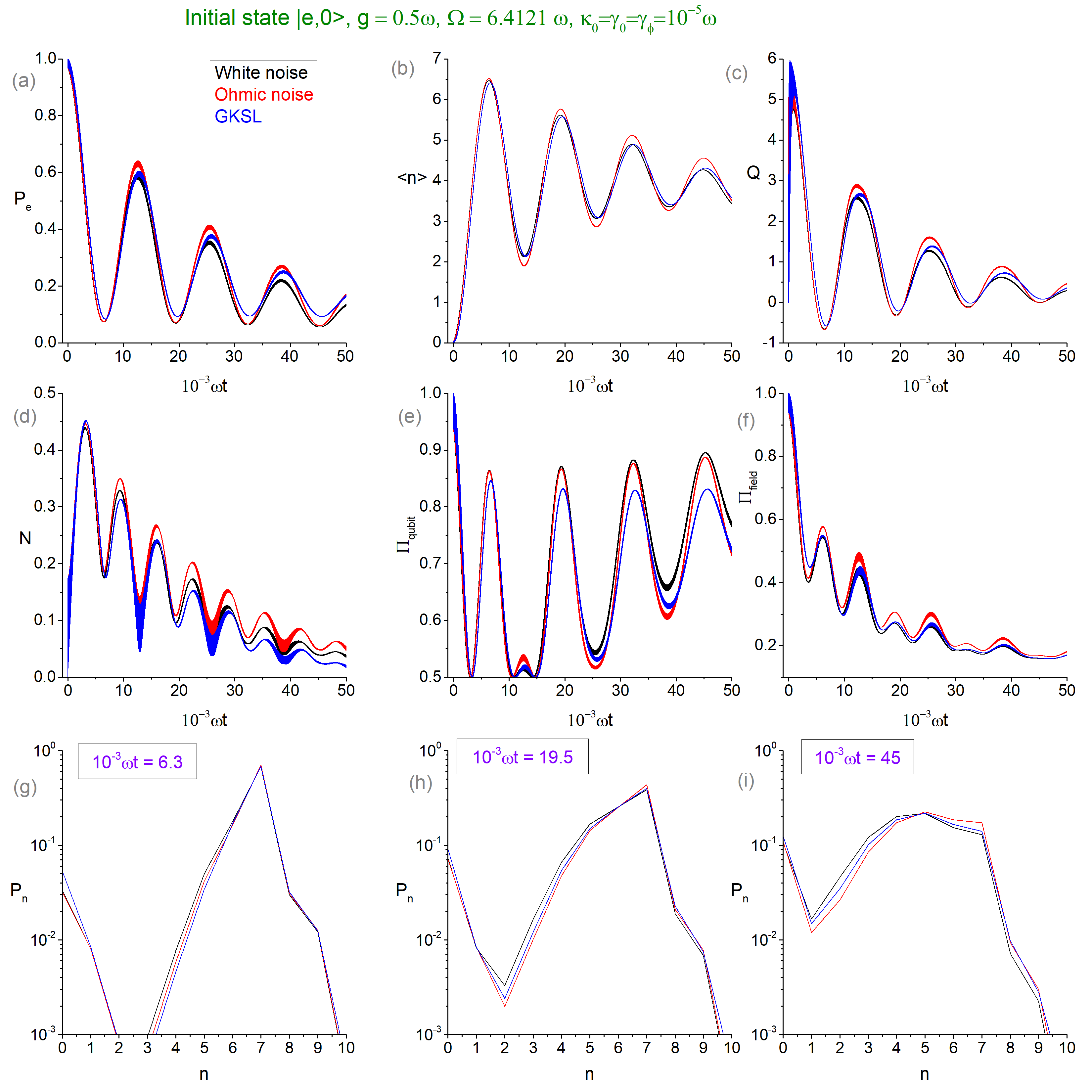} {}
\end{center}
\caption{Similar to Fig. \protect\ref{dodon.fig15}, but for the seven-photon
light--matter resonance: $g=0.5\protect\omega $ and $\Omega =6.4121\protect%
\omega $.}
\label{dodon.fig16}
\end{figure}

Figures \ref{dodon.fig15} and \ref{dodon.fig16} illustrate the system
behavior for the 5- and 7-photon resonances (for the state $|e,0\rangle $)
and larger light-matter coupling strengths. We maintained the same
dissipative rates as in Figure \ref{dodon.fig14} and set the other
parameters as: $g=0.3\omega $, $\Omega =4.7736\omega $ (5-photon resonance)
and $g=0.5\omega $, $\Omega =6.4121\omega $ (7-photon resonance). As
expected, the discrepancies in the predictions of different dissipative
models become more pronounced with the increase of $g$, although the
qualitative behavior remains similar.

\section[Nonstationary Rabi Hamiltonian with post-selection]{Nonstationary Rabi Hamiltonian with\\ post-selection\label{dodon.post}}

Finally, we consider photon generation from the initial vacuum state $%
|g,0\rangle$ induced by modulation of the system parameters, in analogy with
the dynamical Casimir effect. We assume a modulation of the qubit transition
frequency,
\begin{equation}
\Omega (t)=\Omega _{0}\left\{ 1+\varepsilon \sin [\eta (t)t]\right\} ~,
\label{dodon.td}
\end{equation}
where $\Omega _{0}$ is the bare qubit frequency, $\varepsilon \ll \Omega
_{0} $ is the relative modulation amplitude, and $\eta (t)$ is the
time-dependent modulation frequency. Following the proposal of Ref.~\cite%
{dodon.schelb}, we consider a slow linear sweep, $\eta (t)=\eta _{0}+\alpha t
$, where $\eta _{0} $ and $\alpha$ are constant parameters.

Our first objective is to compare the predictions of the DME and the GKSL ME
for the parameter regime analyzed in Ref.~\cite{dodon.schelb}, since the DME
was not considered in that work. The second objective is to investigate the
effect of post-selecting the final cavity-field state conditioned on
detecting the qubit in the ground state. In this case, the cavity-field
density operator at time $t$ becomes
\begin{equation}
\rho _{PS}\left( t\right) = \frac{\mathrm{Tr}_{q}\left( |g\rangle \langle
g|\rho \left( t\right) \right)} {\mathrm{Tr}\left( |g\rangle \langle g|\rho
\left( t\right) \right)} ~,
\end{equation}
where $\rho \left( t\right)$ is the total density operator of the
qubit--cavity system.

Ref.~\cite{dodon.schelb} showed that the generated cavity-field states may
find applications in quantum metrology. Without entering into detailed
derivations, we evaluate here the quantity known as the \emph{quantum Fisher
information} (QFI) for phase estimation under unitary encoding. As discussed
in several reviews~\cite{dodon.re1,dodon.re2,dodon.avs,dodon.Sidhu}, it is
given by
\begin{equation}
\mathcal{F}_{ph}=\frac{1}{2}\sum_{i,j} \frac{\left( p_{i}-p_{j}\right) ^{2}%
} {p_{i}+p_{j}} \left\vert \langle i|\hat{n}|j\rangle \right\vert ^{2}\,,
\label{dodon.QFI}
\end{equation}
where $p_{i}$ and $|i\rangle$ are the eigenvalues and eigenvectors of $\rho
_{PS}\left( t\right)$. Any quantum state satisfying $\mathcal{F}%
_{ph}>\langle n\rangle$, where $\langle n\rangle$ denotes the mean photon
number of the state, possesses metrological advantage over any classical
state of light.

In addition, for the estimation of a \emph{phase-space displacement} of a
single-mode field, one must consider the $2\times 2$ quantum Fisher
information matrix with elements
\begin{equation}
\left( F_{disp}\right) _{kl}= \sum_{i,j} \frac{\left( p_{i}-p_{j}\right) ^{2}%
} {p_{i}+p_{j}} \langle i|x^{\left( k\right) }|j\rangle \langle j|x^{\left(
l\right) }|i\rangle ~,
\end{equation}
where $x^{\left( 1\right) }=a+a^{\dagger }$ and $x^{\left( 2\right) }=\left(
a-a^{\dagger }\right) /i$. The \emph{average Fisher information} over all
quadrature directions is
\begin{equation}
M_{av}=\frac{1}{4}\mathrm{Tr}\left( \mathbf{F}_{disp}\right) ~,  \label{dodon.Mav}
\end{equation}
while the \emph{maximum Fisher information} is
\begin{equation}
M_{opt}=\frac{1}{2}\lambda _{\max }\left( \mathbf{F}_{disp}\right) ~,
\label{dodon.Mopt}
\end{equation}
where $\lambda _{\max }\left( \mathbf{F}_{disp}\right)$ denotes the largest
eigenvalue of $\mathbf{F}_{disp}$. Any quantum state satisfying $M_{av}(\rho
_{PS})>1$ or $M_{opt}(\rho _{PS})>1$ exhibits metrological advantage for
displacement estimation compared to classical states of light.

\begin{figure}[h]
\begin{center}
\includegraphics[width=1.03\textwidth]{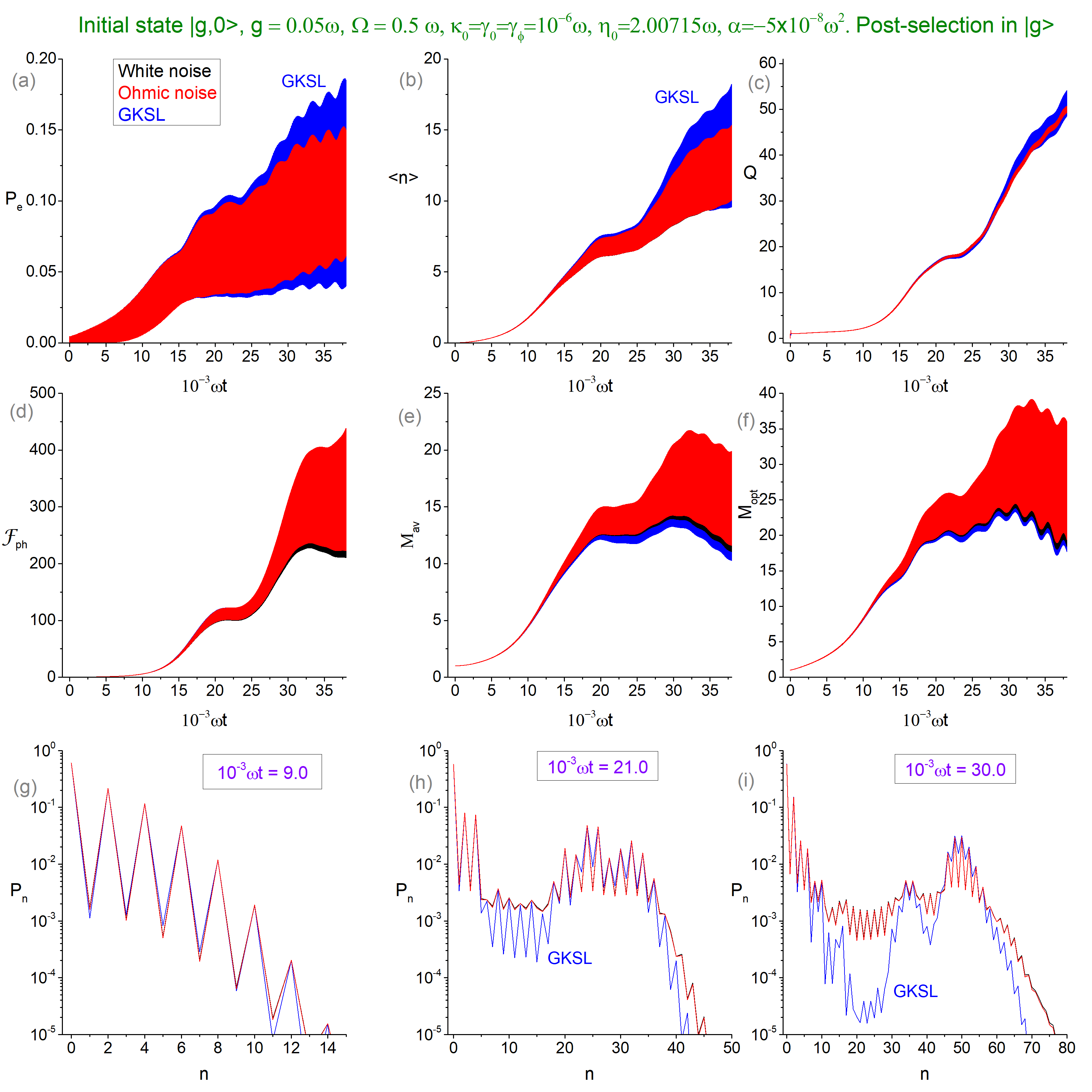} {}
\end{center}
\caption{Nonstationary Rabi Hamiltonian with time-dependent qubit frequency,
Eq. (\protect\ref{dodon.td}). a) Qubit excitation probability. b-f) Behavior of
the average photon number, Mandel $Q$-factor, $\mathcal{F}_{ph}$, $M_{av}$
and $M_{opt}$ for the post-selected cavity field state $\protect\rho %
_{PS}\left( t\right) $, which is projected if the qubit is found in the
ground state at the time $t$ (with the corresponding probability $%
P_{g}=1-P_{e}$). g-i) Photon number probability distribution
at some time instants. Parameters: $g=0.05\protect\nu $, $\Omega _{0}=0.5%
\protect\omega $, $\protect\varepsilon _{\Omega }=0.08$, $\protect\gamma %
_{0}=\protect\gamma _{\protect\phi }=\protect\kappa _{0}=10^{-6}\protect%
\omega $, $\protect\eta _{0}=2.00715\protect\omega $ and $\protect\alpha %
=-5\times 10^{-8}\protect\omega ^{2}$. The nonzero probabilities of odd
photon numbers are due to the dissipation.}
\label{dodonov.fig17}
\end{figure}

\begin{figure}[h]
\begin{center}
\includegraphics[width=1.03\textwidth]{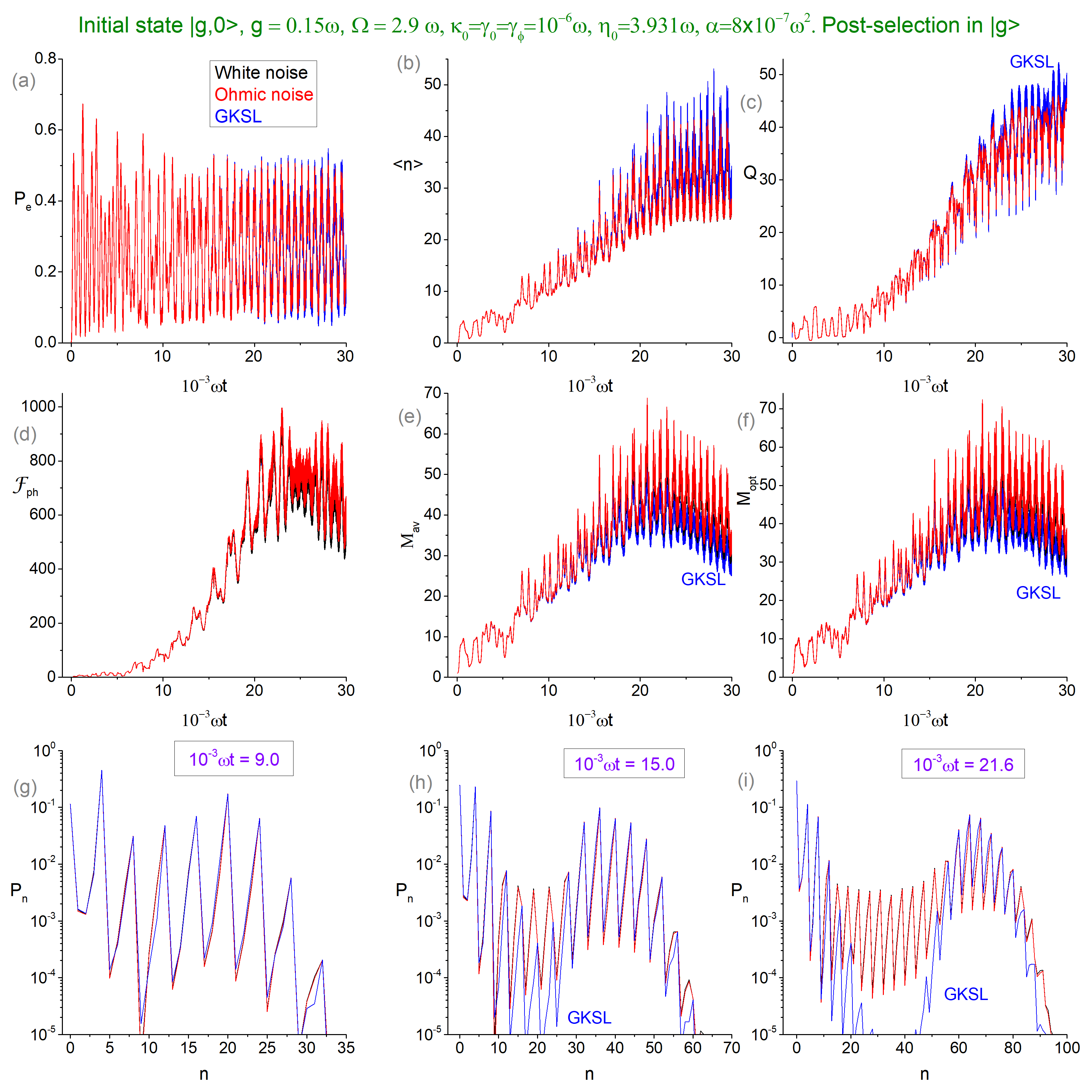} {}
\end{center}
\caption{Similar to Fig. \protect\ref{dodonov.fig17} but for parameters $%
g=0.15\protect\omega $, $\Omega =2.9\protect\omega $, $\protect\eta %
_{0}=3.931\protect\omega $ and $\protect\alpha =8\times 10^{-7}\protect%
\omega ^{2}$. This modulation frequency induces 4-photon transitions between
the system dressed states, as attested by the panels (g)-(i).}
\label{dodonov.fig18}
\end{figure}

We solved numerically the DME and the GKSL ME for the parameters used in
Ref.~\cite{dodon.schelb} and found that for initial times, $\omega t\lesssim2\times 10^4$, the predictions of the three
dissipative models are nearly indistinguishable (data not shown for
brevity), thereby supporting the validity of the results reported in Ref.~%
\cite{dodon.schelb}, where only the GKSL ME was employed. To present new
results and highlight differences between the dissipative approaches, panels~%
\ref{dodonov.fig17}a and \ref{dodonov.fig18}a display the excited-state
probability of the qubit for modulations inducing two- and four-photon
transitions, respectively. Since $P_{g}\gtrsim 0.5$ at all times,
post-selection of the cavity field conditioned on detecting the qubit in the
ground state $|g\rangle $ remains experimentally feasible. We therefore
analyze the properties of the post-selected cavity state.

Specifically, we evaluate the mean photon number, the Mandel $Q$-factor, the metrological quantities~(\ref{dodon.QFI}), (\ref{dodon.Mav}), and (\ref{dodon.Mopt}), as well as the photon-number probability distribution at selected time instants. For early times, $\omega t \lesssim 2\times 10^4$, the predictions of the three dissipative models are nearly identical in these weak-dissipation regimes, even for ultrastrong coupling $g = 0.15\omega$. At later times, the predictions of the DME remain very similar for white and Ohmic reservoirs, but start to deviate quantitatively from those of the GKSL master equation. Furthermore, at longer times, the post-selected cavity-field states exhibit a clear metrological advantage and feature a photon-number distribution with a highly nontrivial, nonmonotonic structure, including a pronounced tail at large $n$.


\section{Summary\label{dodon.summary}}

In this chapter, we have presented a detailed numerical investigation of the
dissipative dynamics of the quantum Rabi model in the ultrastrong coupling
regime, considering several paradigmatic states of quantum optics: coherent,
Schr\"{o}dinger cat, squeezed, thermal, and vacuum states. We explored
different detunings and coupling strengths in the range $0.05\omega $ to $%
0.8\omega $, including multiphoton resonance regimes. The analysis focused
on the time evolution of key observables such as the atomic excitation
probability, mean photon number, Mandel $Q$-factor, negativity, purities,
and photon-number probability distributions. We also examined the regime of
external time-dependent modulation of the qubit parameters \cite%
{dodon.schelb}, where multiphoton processes enable photon generation from
the vacuum via two- and four-photon transitions. In this context, we
demonstrated that postselection protocols can be employed to engineer and
tailor the properties of the cavity field.

Our results reveal that, in certain parameter regimes, the predictions of
the standard GKSL master equation (ME) and the dressed master equation (DME)
differ substantially across all observables considered. In other regimes,
however, the two approaches lead to qualitatively similar dynamics for some
quantities while exhibiting noticeable quantitative differences for others.
Consistent with previous observations~\cite{dodon.Settineri}, we find that
in the dispersive regime the DME tends to predict stronger coherence losses:
high-frequency oscillations superimposed on the primary dynamical envelope
are more heavily damped, and quantities such as purities and negativity
decay faster than in the GKSL treatment. Nevertheless, without resorting to
the generalized dressed master equation, it is not possible to state
definitively that such an overestimation occurs universally.

The main conclusion of this work is therefore methodological as well as
physical. When investigating new parameter regimes of the
ultrastrong-coupling Rabi model, both the GKSL ME and the DME should be
solved numerically and compared carefully, possibly using computational
strategies such as those illustrated here. At the same time, for the
parameter ranges analyzed in this chapter, our extensive numerical results
provide a practical reference for anticipating both the qualitative trends
and quantitative deviations associated with each master-equation approach.

\section{Acknowledgments}

A.P.C acknowledges the ﬁnancial support by the Brazilian agency Coordenação de Aperfeiçoamento de Pessoal de Nível Superior (CAPES, Finance Code 001). H.S.R.O. acknowledges the financial support of the
Brazilian agency Funda\c{c}\~{a}o de Apoio \`{a} Pesquisa do Distrito
Federal (FAPDF, program PIBIC). A. D. acknowledges a partial financial support of the Brazilian agency Funda%
\c{c}\~{a}o de Apoio \`{a} Pesquisa do Distrito Federal (FAPDF, grant number
00193-00001817/2023-43).

\renewcommand{\bibname}{References} \begingroup
\let\cleardoublepage\relax

\endgroup

\end{document}